\documentclass{article}

\usepackage[dvipsnames]{xcolor}
\usepackage[colorlinks=true,linkcolor=blue!45!black,citecolor=blue!45!black,urlcolor=blue!45!black]{hyperref}
\usepackage{graphicx,float,multicol}
\usepackage[toc,page]{appendix}
\usepackage{bookmark}
\usepackage{amsmath,amsbsy,bm,amssymb}
\usepackage{cleveref}
\usepackage{algorithm}
\usepackage{algpseudocode}
\usepackage{etoolbox}
\usepackage{tikz}
\usepackage{enumerate}
\usepackage[symbol]{footmisc}
\usepackage[group-minimum-digits=3]{siunitx}

\DeclareMathOperator*{\argmax}{arg\,max}
\DeclareMathOperator*{\argmin}{arg\,min}
\newrobustcmd*{\plotsquare}[2]{\tikz{\filldraw[draw=#1,fill=#1,#2] (0,0) rectangle (.2cm,.2cm);}}
\newrobustcmd*{\plotcircle}[1]{\tikz{\filldraw[draw=#1,fill=#1] (0,0) circle [radius=.1cm];}}
\newrobustcmd*{\plottriangle}[1]{\tikz{\filldraw[draw=#1,fill=#1] (0,0) -- (.2cm,0) -- (.1cm,.2cm);}}
\newrobustcmd*{\plotline}[1]{\tikz{\draw[very thick,#1] (0,.1cm) -- (.4cm,.1cm);
 \draw[draw=none, use as bounding box](0,0) rectangle (.2cm,.2cm);}}
\crefname{figure}{Figure}{Figures}
\crefname{tabular}{Table}{Tables}
\crefname{algorithm}{Algorithm}{Algorithms}
\crefname{equation}{Eq.}{Eqs.}
\crefname{appendix}{}{}
\Crefname{figure}{Figure}{Figures}
\Crefname{tabular}{Table}{Tables}
\Crefname{algorithm}{Algorithm}{Algorithms}
\Crefname{equation}{Eq.}{Eqs.}
\Crefname{appendix}{}{}

\begin{document}

\title{Optimized data exploration applied to the simulation of a chemical process}
\date{}
\author{}
\maketitle

\vspace{-.5cm}
\begin{center}
Raoul Heese\textsuperscript{a,b,}\footnote[1]{Corresponding author: \href{mailto:raoul.heese@itwm.fraunhofer.de}{raoul.heese@itwm.fraunhofer.de}}, Micha{\l} Walczak\textsuperscript{a,b}, Tobias Seidel\textsuperscript{b},\\Norbert Asprion\textsuperscript{c}, Michael Bortz\textsuperscript{a,b}\\
\vspace{.5cm}
\textit{\textsuperscript{a}Fraunhofer Center for Machine Learning\\
\textsuperscript{b}Fraunhofer ITWM Optimization Department, Fraunhofer-Platz 1, 67663 Kaiserslautern, Germany\\
\textsuperscript{c}Chemical and Process Engineering BASF SE, Carl-Bosch-Str. 38, 67056 Ludwigshafen, Germany}
\end{center}

\vspace{1cm}
\begin{center}
\textbf{Abstract}
\end{center}
In complex simulation environments, certain parameter space regions may result in non-convergent or unphysical outcomes. All parameters can therefore be labeled with a binary class describing whether or not they lead to valid results. In general, it can be very difficult to determine feasible parameter regions, especially without previous knowledge. We propose a novel algorithm to explore such an unknown parameter space and improve its feasibility classification in an iterative way. Moreover, we include an additional optimization target in the algorithm to guide the exploration towards regions of interest and to improve the classification therein. In our method we make use of well-established concepts from the field of machine learning like kernel support vector machines and kernel ridge regression. From a comparison with a Kriging-based exploration approach based on recently published results we can show the advantages of our algorithm in a binary feasibility classification scenario with a discrete feasibility constraint violation. In this context, we also propose an improvement of the Kriging-based exploration approach. We apply our novel method to a fully realistic, industrially relevant chemical process simulation to demonstrate its practical usability and find a comparably good approximation of the data space topology from relatively few data points.

%\author[addresscit,addressitwm]{Raoul Heese\corref{mycorrespondingauthor}}
%\ead{raoul.heese@fraunhofer.itwm.de}
%\author[addresscit,addressitwm]{Micha{\l} Walczak}
%\author[addressitwm]{Tobias Seidel}
%\author[addressbasf]{Norbert Asprion}
%\author[addresscit,addressitwm]{Michael Bortz}
%
%\address[addresscit]{Fraunhofer Center for Machine Learning}
%\address[addressitwm]{Fraunhofer ITWM Optimization Department, Fraunhofer-Platz 1, 67663 Kaiserslautern, Germany}
%\address[addressbasf]{Chemical and Process Engineering BASF SE, Carl-Bosch-Str. 38, 67056 Ludwigshafen, Germany}

\newpage
%%%%%%%%%%%%%%%%%%%%%%%%%%%%%%%%%%%%%%%%%%%%%%%%%%%%%%%%%%% body.tex 
\section{Introduction} \label{sec:Introduction}
Chemical process design is in its nature a multicriteria optimization (MCO) task~\cite{Miettinen1999,Bortz2014}. To find a Pareto-optimal design~\cite{Geoffrion1968}, an engineer needs to explore multiple trade-offs between competing (design) parameters and objectives. Recently, powerful decision support tools have been developed for usage in industrial contexts to aid an engineer in this design process. These tools enable him to navigate through Pareto-efficient solutions from simulation runs performed for different parameter variations~\cite{Bortz2014,Burger2014}.\par
In a typical industrial application, a flowsheet simulation of a chemical process involves solving several hundreds to several thousands of nonlinear equations for each specified parameter combination. Only certain combinations of the parameters lead to numerically converging, physically reasonable results. Therefore, in order to estimate the Pareto frontier it is necessary to determine valid parameter combinations, i.\,e., to find the operation window or feasibility region in the parameter space.\par
A binary feasibility classification task in this context can be challenging due to an often vast parameter space and the computational effort involved in solving the system of nonlinear equations. There exist different approaches from various branches of engineering how to solve this problem. Many studies propose to reduce the computational cost by approximating the feasible region. For example, in Ref.~\cite{Jeong2012} parameters are classified according to their feasibility using a support vector machine (SVM)~\cite{Cortes1995} combined with k-means clustering to reduce the model training cost.\par
An alternative to binary classification of feasible points is to train a surrogate model of a continuous feasibility function, which reflects by how much the constraints which define the feasibility range are violated. The surrogate model is usually trained adaptively by sampling regions with high model uncertainty and the vicinity of the predicted boundary. Suggested ways to construct a surrogate model include high dimensional model representation techniques~\cite{Banerjee2010}, Kriging-based modelling~\cite{Boukouvala2012,Boukouvala2014}, and most recently, radial basis function (RBF)-based modelling~\cite{Wang2017}. The RBF-based modeling outperforms Kriging-based modeling in terms of achieving better accuracy with fewer sampling points, however, both approaches exhibit limitations in five- and higher-dimensional problems~\cite{Wang2017}. Moreover, in certain applications it is impossible to access the information about maximum constraints violation.\par
In this manuscript, we present a novel method to sequentially explore feasibility classifications. Inspired by Ref.~\cite{Boukouvala2014} our proposed algorithm does not only provide us with an adaptive estimator for feasibility, but can also take an optimization target into account. This allows us to focus the exploration on parameter regions of interest. We therefore consider our approach as a method for optimized data exploration, which exceeds bare feasibility classification. Briefly put, we propose an adaptive screening of an operation window combining feasibility classification and optimization.\par
For demonstration purposes we apply our method to the simulation of a realistic chemical process. In particular, we consider the simulation of two serialized distillation columns to separate an azeotropic mixture of chloroform and acetone~\cite{Biegler1997}. We show that we can improve the estimation of feasible regions in comparison with a uniform grid approach and a latin hypercube sampling (LHS)~\cite{McKay1979} approach. We use \emph{Chemasim}, a BASF in-house flowsheet simulator, to run the simulations. By design, Chemasim, like other flowsheet simulators, does not provide a complete quantification of equation violations for unfeasible parameters, hence a surrogate model of a continuous feasibility function can not be applied. Our algorithm, however, is based entirely on a binary feasibility classification, which we can directly extract from Chemasim.\par
In the following, we will first present a formal description of our algorithm. Subsequently, we will explain its functionality with the help of a toy example. Using a benchmark, we will show the strengths of our method in comparison with a Kriging-based exploration approach for a binary feasibility classification scenario. We also propose an improvement of this Kriging-based approach for a discrete feasibility constraint violation. Finally, we will outline the chemical process simulation to which we have applied our algorithm and present the results. We will conclude with a brief summary and outlook.

\section{Optimized data exploration} \label{sec:Optimized data exploration}
Our method of optimized data exploration can be considered as a sequential design of experiments, where each new experiment (i.\,e., each new simulation evaluation) is chosen based on a utility function. Summarized, the method consists of four major steps:
\begin{enumerate}[(i)]
\item \textit{Start-up}: Ensure that an initial data set is available. This preliminary step is required to ensure at least basic knowledge about the data topology. Therefore, one may use previously obtained data or evaluate a set of parameters, e.\,g., on a regular grid.
\item \textit{Choice}: Choose the parameter for which the utility function is maximized. The utility function makes use of the previously obtained data to interpolate or extrapolate missing information.
\item \textit{Evaluation}: Evaluate the simulation for the chosen parameter.
\item \textit{Repetition}: Repeat from (ii) or finish the exploration based on a termination criterion.
\end{enumerate}
The choice of a suitable utility function and a sufficient initial knowledge about the data are of course crucial for the success of our approach.\par
From a practical perspective, data exploration is often driven by an optimization, hence we consider in the following that we will not only evaluate a simulation outcome for its validity but will also extract some kind of quantitative optimization target from the result. In order to emphasize on the main aspects of our data exploration method, we will only consider a single optimization target. A MCO problem can, however, always be reduced to this one-dimensional case by an appropriate scalarization of its targets.\par
In this section, we will first describe the general framework of our method in which we introduce the main formal ingredients. We will subsequently explain the exploration algorithm itself in more detail. A toy example will help us to demonstrate our method.

\subsection{Framework} \label{sec:Framework}
We use simulation evaluations and estimators to generate data. Therefore, we will first formally define those concepts. Furthermore, we will discuss our choice of a utility function. These definitions will serve as a framework for the following studies.

\subsubsection{Data generation}
We consider a simulation $\mathcal{S}$ which is described by a mapping
\begin{align} \label{eqn:Smap}
\mathcal{S} : \boldsymbol{\chi} \rightarrow S
\end{align}
of parameters $\mathbf{x}$ from the compact $p$-dimensional parameter space $\boldsymbol{\chi} \subset \mathbb{R}^p$ onto the solution space
\begin{align} \label{eqn:S}
S \equiv \eta \otimes \tau.
\end{align}
This solution space consists of two parts. First, a classification space 
\begin{align} \label{eqn:eta}
\eta \equiv \{\mathrm{valid},\mathrm{invalid}\}
\end{align}
containing two classes which describe whether the simulation outcome $y \equiv y(\mathcal{S},\mathbf{x}) \in \eta$ was valid (i.\,e., numerically convergent and physically reasonable in the sense that the numerically obtained simulation result fulfills a set of predefined conditions) or invalid. And second, the optimization target space $\tau \subset \mathbb{R}$ which contains all possible results for the optimization target $t \equiv t(\mathcal{S},\mathbf{x}) \in \tau$. The optimization target is meaningful only for valid simulation outcomes. Without loss of generality we assume in the following that a maximization of $t$ is considered optimal. The symbol $\otimes$ in~\cref{eqn:S} denotes a tensor product.\par
Each evaluation of the simulation therefore leads to a data point
\begin{align} \label{eqn:d}
d(\mathbf{x}) \equiv (\mathbf{x},y,t)
\end{align}
given by the evaluated parameter $\mathbf{x}$, the corresponding outcome $y$ and the optimization target $t$ from the mapping $\mathcal{S}$, \cref{eqn:Smap}. Although the optimization target is meaningless for invalid outcomes, we still include it in $d(\mathbf{x})$ to achieve a unified notation for valid and invalid data points. Note that we assume that the result of a simulation is purely deterministic and completely defined by the choice of parameters $\mathbf{x}$. The collection of results from $n$ evaluations consequently allows us to define a data set
\begin{align}
D \equiv D(\mathbf{x}_1,\dots,\mathbf{x}_n) \equiv \{d(\mathbf{x}_1),\dots,d(\mathbf{x}_n)\}.
\end{align}
Data exploration is achieved by adding new elements to such a set. For reasons of convenience we use
\begin{align} \label{eqn:Dx}
D_x \equiv \{ \mathbf{x}_1,\dots,\mathbf{x}_n \}
\end{align}
to denote the corresponding collection of evaluated parameters.

\subsubsection{Prediction} \label{sec:Framework:prediction}
A sufficiently large data set allows us to perform model predictions for solutions of yet unevaluated parameters with the help of an estimator
\begin{align} \label{eqn:Emap}
\mathcal{E} : \mathcal{D} \otimes \boldsymbol{\chi} \rightarrow S \otimes \eta_p
\end{align}
which maps from the space of possible data sets $\mathcal{D}$ and the parameter space $\boldsymbol{\chi}$ onto the solution space $S$, \cref{eqn:S}, and the predicted outcome probability space $\eta_p \equiv [0,1]$. The solution space contains the predicted outcome $\hat{y} \equiv \hat{y}(\mathcal{E},D,\mathbf{x}) \in \eta$ and the predicted optimization target $\hat{t} \equiv \hat{t}(\mathcal{E},D,\mathbf{x}) \in \tau$, whereas the predicted outcome probability space contains the probability of the predicted outcomes $\hat{p}_{\hat{y}} \equiv \hat{p}_{\hat{y}}(\mathcal{E},D,\mathbf{x}) \in [0,1]$. Briefly put, $\mathcal{E}$ allows us to predict simulation outcomes with their associated probability and optimization targets for arbitrary parameters based on previous simulation results.\par
So far, $\mathcal{E}$ is considered completely general and our algorithm is not limited to a specific choice. However, it has turned out in practice that $\mathcal{E}$ is best represented by two independent estimators: First, a kernel SVM used for a classification of the outcome $\hat{y}$ and second, a kernel ridge regression (RR)~\cite{Murphy2012} to predict the optimization target $\hat{t}$. The probability of the predicted outcomes $\hat{p}_{\hat{y}}$ is then obtained from Platt scaling~\cite{Platt1999}. The kernel RR can be understood as a surrogate model for the optimization target. For both of these estimations we use the well-known kernel method~\cite{Hastie2013}, which is discussed in \cref{app:kernel methods} in more detail.

\subsubsection{Utility function} \label{sec:Framework:utility function}
At the heart of our data exploration method lies the utility function
\begin{align} \label{eqn:U}
U(\mathcal{E},D,\mathbf{x},\mathbf{w}) & \equiv \frac{\mathbf{w}^T \mathbf{u}(\mathcal{E},D,\mathbf{x})}{||\mathbf{w}||_1}
\end{align}
which describes the estimated exploration benefit of evaluating the parameter $\mathbf{x}$ based on a previously obtained data set $D$. Therefore, in each sequential exploration step a new evaluation of the simulation is chosen for the parameter
\begin{align} \label{eqn:xnew}
\textbf{x}_{\mathrm{new}} = \argmax_{\textbf{x} \in \boldsymbol{\chi}} U(\mathcal{E},D,\mathbf{x},\mathbf{w})
\end{align}
with the best utility score. The two ingredients of the utility function are the utility vector
\begin{align} \label{eqn:u}
\mathbf{u} \equiv \mathbf{u}(\mathcal{E},D,\mathbf{x}) \equiv \begin{pmatrix}U_s(\mathcal{E},D,\mathbf{x}) \\ U_o(\mathcal{E},D,\mathbf{x}) \\ U_r(D,\mathbf{x})\end{pmatrix}
\end{align}
and the weight vector
\begin{align} \label{eqn:w}
\mathbf{w} \equiv \mathbf{w}(s,o,r) \equiv \begin{pmatrix}s \\ o \\ r\end{pmatrix},
\end{align}
which both consist of three components. Each component of the utility vector can be assigned a straightforward interpretation:
\begin{align} \label{eqn:u:interpretation}
\mathbf{u} \sim \begin{pmatrix}\text{Outcome prediction uncertainty}\\\text{Optimization target prediction}\\\text{Distance to nearest neighbor}\end{pmatrix}.
\end{align}
The components (or weights) $s \geq 0$, $o \geq 0$ and $r \geq 0$ of the weight vector determine the influence of $U_s(\mathcal{E},D,\mathbf{x})$, $U_o(\mathcal{E},D,\mathbf{x})$ and $U_r(D,\mathbf{x})$, respectively, on the utility function. In other words, $\mathbf{w}$ determines the explorative behavior. The expression $||\mathbf{w}||_1 = s+o+r$ in \cref{eqn:U} represents the 1-norm of $\mathbf{w}$ and since by definition $\mathbf{u} \in [0,1]^3$, as we will see below, one has $U(\mathcal{E},D,\mathbf{x},\mathbf{w}) \in [0,1]$. In the following, we will formally define the components of $\mathbf{u}$, \cref{eqn:u}.\par
The first component
\begin{align}
U_s(\mathcal{E},D,\mathbf{x}) & \equiv S(\hat{p}_{\hat{y}}(\mathcal{E},D,\mathbf{x}))
\end{align}
represents the estimated outcome prediction uncertainty based on the Shannon information entropy~\cite{Shannon1948}
\begin{align}
S(\hat{p}_{\hat{y}}) & \equiv - \frac{\hat{p}_{\hat{y}} \ln \hat{p}_{\hat{y}} + ( 1 - \hat{p}_{\hat{y}} ) \ln( 1 - \hat{p}_{\hat{y}} )}{\ln 2}
\end{align}
in bits. Here we have recalled the estimated probability of predicting a valid or invalid outcome $\hat{p}_{\hat{y}}(\mathcal{E},D,\mathbf{x})$ from the estimator mapping, \cref{eqn:Emap}.\par
The second component
\begin{align}
U_o(\mathcal{E},D,\mathbf{x}) & \equiv \begin{cases} 1 & \text{if} \ \hat{t}_{\mathrm{r}}(\mathcal{E},D,\mathbf{x}) > 1 \\ 0 & \text{if} \ \hat{t}_{\mathrm{r}}(\mathcal{E},D,\mathbf{x}) < 0 \\ t_{\mathrm{r}}(\mathcal{E},D,\mathbf{x}) & \text{otherwise} \end{cases}
\end{align}
represents the estimated optimization score. It is based on
\begin{align}
\hat{t}_{\mathrm{r}}(\mathcal{E},D,\mathbf{x}) & \equiv \begin{cases} 0 & \text{if} \ \hat{t}(\mathcal{E},D,\mathbf{x}) < t_{\mathrm{min}}(D) \\ 1 & \text{if} \ \hat{t}(\mathcal{E},D,\mathbf{x}) >  t_{\mathrm{max}}(D) \\ \frac{\hat{t}(\mathcal{E},D,\mathbf{x}) - t_{\mathrm{min}}(D)}{t_{\mathrm{max}}(D) - t_{\mathrm{min}}(D)} & \text{else} \end{cases},
\end{align}
which makes use of the extremal optimization targets
\begin{subequations} \label{eqn:textremal}
\begin{align}
t_{\mathrm{max}}(D) & \equiv \underset{\substack{(\mathbf{x}, y, t) \in D \\ \text{s.\,t.} \ \ y = \text{valid}}}{\max} t
\end{align}
and
\begin{align} \label{eqn:textremal:min}
t_{\mathrm{min}}(D) & \equiv \underset{\substack{(\mathbf{x}, y, t) \in D \\ \text{s.\,t.} \ \ y = \text{valid}}}{\min} t,
\end{align}
\end{subequations}
respectively, to rescale the predicted optimization target $\hat{t}(\mathcal{E},D,\mathbf{x})$, from \cref{eqn:Emap}, in such a way that $\hat{t}_{\mathrm{r}}(\mathcal{E},D,\mathbf{x}) \in [0,1]$.\par
As a last component, the utility vector contains the classification feature space distance~\cite{Scholkopf2000}
\begin{align} \label{eqn:Ur}
U_r(D,\mathbf{x}) & = 1 - \exp ( - \gamma_{\mathrm{C}} \min_{\mathbf{x}^{\prime} \in D_x} || \mathbf{x} - \mathbf{x}^{\prime} ||_2^2 ).
\end{align}
Here, the expression $|| \cdot ||_2$ stands for the 2-norm distance and $\gamma_{\mathrm{C}}$ represents a hyperparameter of the kernel SVM estimator. A derivation of \cref{eqn:Ur} can be found in \cref{app:classification feature space distance}.\par
By definition, $U_r(D,\mathbf{x})$ becomes smaller the more similar the parameter $\mathbf{x}$ is to its nearest neighbor in $D_x$, \cref{eqn:Dx}. Therefore, this expression can also be understood as an artificial repulsion of data points which ensures that newly suggested parameters $\textbf{x}_{\mathrm{new}}$ explore unknown regions of the parameter space $\boldsymbol{\chi}$. The amount of utility reduction with decreasing distance is controlled by $\gamma_{\mathrm{C}}$. In \cref{app:kernel methods} this hyperparameter is discussed in more detail. Note that it is determined during the training of $\mathcal{E}$, \cref{eqn:Emap}, as explained further below.\par
Summarized, we have introduced the utility function $U(\mathcal{E},D,\mathbf{x},\mathbf{w})$, \cref{eqn:U}, based on the utility vector $\mathbf{u}$, \cref{eqn:u}, and the weight vector $\mathbf{w}$, \cref{eqn:w}. The utility vector consists of three components with a distinct meaning, \cref{eqn:u:interpretation}. These components are weighted by the weights $s$, $o$ and $r$ contained in the weight vector. Consequently, we can control the explorative behavior by tuning the weights. 

\subsection{Algorithm} \label{sec:Algorithm}
Our data exploration method is outlined in \cref{alg:doe}. As described in the introduction of this section, it can be partitioned in four major steps. These steps correspond to the following lines of the algorithm:
\begin{enumerate}[(i)]
\item \textit{Start-up}: Line 2 to 11
\item \textit{Choice}: Line 13
\item \textit{Evaluation}: Line 14
\item \textit{Repetition}: Line 12 and 16
\end{enumerate}
As a termination criterion we use a desired number of newly evaluated points $N$. We assume here that the simulation $\mathcal{S}$, \cref{eqn:Smap}, is defined by the application and is not modified during the exploration process. By contrast, the estimator $\mathcal{E}$, \cref{eqn:Emap}, is retrained in each iteration step. For the sake of simplicity we omit $\mathcal{S}$ and $\mathcal{E}$ in the notation.

\begin{algorithm}
\caption{Outline of our novel data exploration algorithm. A detailed description can be found in \cref{sec:Algorithm}.}\label{alg:doe}
\begin{algorithmic}[1]
\Function{Exploration}{$D_{\mathrm{init}},\boldsymbol{\chi},G,\boldsymbol{\chi}_{\mathrm{G}},N,\mathbf{w}$}
\State $D_{\mathrm{expl}} \gets D_{\mathrm{init}}$
\If {$D_{\mathrm{expl}} = \{\}$}
    \State $\mathbf{x}_0,\dots,\mathbf{x}_{G^p-1} \gets \Call{Grid}{G, \boldsymbol{\chi}_{\mathrm{G}}}$
    \ForAll{$\mathbf{x} \in \{\mathbf{x}_0,\dots,\mathbf{x}_{G^p-1}\}$}
    	\State $D_{\mathrm{expl}} \gets D_{\mathrm{expl}} \cup \Call{Simulation}{\mathbf{x}}$
    \EndFor
    \State $n \gets G^p$
\Else
	\State $n \gets 0$
\EndIf
\While{$n < N$}
	\State $\textbf{x} \gets \Call{Suggestion}{D_{\mathrm{expl}},\boldsymbol{\chi},\mathbf{w}}$
	\State $D_{\mathrm{expl}} \gets D_{\mathrm{expl}} \cup \Call{Simulation}{\mathbf{x}}$
	\State $n \gets n + 1$
\EndWhile
\State \Return $D_{\mathrm{expl}}$
\EndFunction
\end{algorithmic}
\end{algorithm}

\subsubsection{Functions}
The entry point to the algorithm is the \textsc{Exploration} function. Its arguments represent an initial data set from previous evaluations $D_{\mathrm{init}}$ (which might also be an empty set), the parameter space to explore $\boldsymbol{\chi}$, the parameter $G \geq 0$ controlling the number of initial evaluations $G^p$ in the hyperrectangular compact set $\boldsymbol{\chi}_{\mathrm{G}} \subseteq \boldsymbol{\chi}$, the desired total number of points to evaluate $N \geq G^p$ and the weight vector $\mathbf{w}$, \cref{eqn:w}, which controls the behavior of the utility function, \cref{eqn:U}. Furthermore, the \textsc{Exploration} function contains three implicit functions:
\begin{itemize}
\item \textsc{Simulation}: Performs the mapping $\mathcal{S}$ and returns the corresponding data point $d(\mathbf{x})$.
\item \textsc{Suggestion}: Performs two steps in order to obtain the next parameter to evaluate. First, the estimator $\mathcal{E}$ for the current data set $D_{\mathrm{expl}}$ is trained. Second, \cref{eqn:xnew} is evaluated and the parameter $\textbf{x}_{\mathrm{new}}$ with the best utility is returned.
\item \textsc{Grid}: Returns a set of $G^p \leq N$ parameters which constitute a regular grid in the parameter subspace $\boldsymbol{\chi}_{\mathrm{G}}$. If previous knowledge about the data topology is available, it is generally reasonable to choose a starting grid in such a way that both valid and invalid solutions are sampled in a region of an expectably good optimization target.
\end{itemize}
After $N$ points have been evaluated, the explored data set $D_{\mathrm{expl}} \supset D$ is returned.\par
The simulation $\mathcal{S}$, \cref{eqn:Smap}, is considered completely general up to this point. Therefore, our algorithm is universal and can be applied in many different scenarios. We will specify simulation mappings further below in explicit examples for data exploration.

\subsubsection{Estimator training}
During the training of $\mathcal{E}$ we tune the hyperparameters of the SVM and the RR independently in each iteration step by cross validation~\cite{Hastie2013} as soon as the size of the data set $D_{\mathrm{expl}}$ allows it. For training and prediction we use standardized parameters by removing the mean and scaling to unit variance. All available data at each iteration step is considered as training data.\par
The optimization target $t$ corresponding to an invalid outcome $y$ is meaningless, but it might nevertheless be of practical use to be able to assign some numerical value to it, e.\,g., to train an estimator. Therefore, we assume in the following that invalid optimization targets taken from a data set $D$ correspond to the worst valid optimization target of this data set $t_{\mathrm{min}}(D)$, \cref{eqn:textremal:min}, for all practical purposes.\par
Moreover, we assume that the initial data set $D_{\mathrm{init}}$ or the data set obtained from the initial grid sampling is sufficiently large so that a suitable estimator of our choice is well-defined. Otherwise, either the start-up step has to be changed accordingly or a different estimator has to be chosen. We will not further discuss such pathological cases.

\section{Demonstration} \label{sec:Demonstration}
To illustrate our data exploration method from the previous section, we will first present a two-dimensional toy example (i.\,e., $p=2$). Specifically, we consider parameters
\begin{align} \label{eqn:toy:x}
\mathbf{x} \equiv \begin{pmatrix} x_1 \\ x_2 \end{pmatrix}
\end{align}
in the parameter space
\begin{align} \label{eqn:toy:chi}
\boldsymbol{\chi}_{\mathrm{toy}} \equiv [-2,2] \otimes [-2,2].
\end{align}
Thus, the parameter space is a simple square of edge length 4.

\subsection{Toy simulation}
The chosen toy simulation $\mathcal{S}_{\mathrm{toy}}$, \cref{eqn:Smap}, is given by
\begin{subequations} \label{eqn:toy}
\begin{align} \label{eqn:toy:y}
y(\mathcal{S}_{\mathrm{toy}},\mathbf{x}) \equiv \begin{cases} \mathrm{valid} & \text{if} \ ||\mathbf{x}||_2 \leq \sqrt{3} \land x_1 x_2 \geq 0 \\ \mathrm{invalid} & \text{otherwise} \end{cases}
\end{align}
and
\begin{align} \label{eqn:toy:t}
t(\mathcal{S}_{\mathrm{toy}},\mathbf{x}) \equiv x_1,
\end{align}
\end{subequations}
respectively, where $||\mathbf{x}||_2$ denotes the 2-norm of $\mathbf{x}$. Thus, simulation outcomes $y$ are considered valid within and on a circle of radius $\sqrt{3}$ in the first and third quadrant and invalid otherwise as shown in \cref{fig:toy-pspace}. This particular example is interesting to study data exploration behavior because it has both sharp and round edges and two distinct feasibility areas only connected by a single point. The optimization target $t$ directly corresponds to $x_1$.\par
We assume no previous knowledge about the data so that $D_{\mathrm{init}} = \{\}$. The initial grid is chosen to expand uniformly across the whole parameter space $\boldsymbol{\chi}_{\mathrm{G}} = \boldsymbol{\chi}_{\mathrm{toy}}$, \cref{eqn:toy:chi}. To solve \cref{eqn:xnew} numerically in each iteration step we use a differential evolution approach~\cite{Storn1997}. Although our proposed algorithm, \cref{alg:doe}, is completely deterministic in its general form, the usage of cross validation for the estimator training, the statistics involved in the Platt scaling and a differential evolution approach to solve the optimization problem introduce a certain degree of randomness. The presented results are therefore chosen as representative examples.

\subsection{Results} \label{sec:Demonstration:Results}
The results are shown in \cref{fig:toy-explcomp}. Each row, (1) to (5), corresponds to a different set of exploration hyperparameters $N$, $G$, $\boldsymbol{\chi}_{\mathrm{G}}$, $s$, $o$ and $r$ as summarized in \cref{tab:toy-setups}. We will hereafter refer to a set of such hyperparameters as a setup. Each setup yields a different explored data set $D_{\mathrm{expl}}$. Column (a) shows the evaluated outcomes $y$, \cref{eqn:toy:y}, and labels them as valid (\plotcircle{green}) or invalid (\plottriangle{red}). The contours (\plotline{blue}) separate the valid from the invalid predicted outcomes $\hat{y}$, \cref{eqn:Emap}, as given by the final estimator. The corresponding regions are shaded accordingly (\plotsquare{green}{opacity=.4} and \plotsquare{red}{opacity=.3}). Column (b) shows the optimization target $t$, \cref{eqn:toy:t}, for valid outcomes as color-interpolated markers from the best possible result $t=\sqrt{3}$ (\plotcircle{cyan}) to the worst possible result $t=-\sqrt{3}$ (\plotcircle{magenta}). Also shown are the invalid outcomes (\plottriangle{red}). For the first and second row we also mark the iteration number $n+1$ from \cref{alg:doe}, i.\,e., the chronological order in which the points have been evaluated after the initial grid had been set up.\par
Apparently, setup (1) already provides us with a rough estimate of the parameter topology with 16 evaluations, which is then refined by setups (2) and (3) with 9 and 39 additional evaluations, respectively. For all of these first three setups, the optimization target has no effect on the utility. This situation changes for setup (4), where we make it the main contribution to the utility function. Consequently, the evaluations are concentrated in an area with high values of $t$ and the lower left part of the topology is hardly explored. Finally, setup (5) shows a pure grid approach in the framework of our algorithm for comparison, which is finished after the initial sampling step since $G^2 = N$.

\subsubsection{Relative success rate and score}
In~\cref{app:relative success rate and score} we define the relative success rate
\begin{align} \label{eqn:R}
R \equiv R(D_{\mathrm{expl}})
\end{align}
and the score
\begin{align} \label{eqn:sigma}
\sigma \equiv \sigma(D_{\mathrm{expl}})
\end{align}
as a quality measure for our estimator. The score represents the fraction of correct outcome predictions performed for all parameters in the explored data set $D_{\mathrm{expl}}$, whereas $R$ is defined as the fraction of correct outcome predictions for almost all parameters in the parameter space $\boldsymbol{\chi}$. In other words, $\sigma$ is a local and $R$ a global quality measure of our estimator.\par
The fraction 
\begin{align} \label{eqn:g}
g \equiv \frac{\sigma}{R}
\end{align}
consequently tells us how well the outcome predictions from the estimator obtained from the explored data set generalize to the whole parameter space. If $D_{\mathrm{expl}}$ is a representative subset of $\boldsymbol{\chi}$, $g$ can consequently be used to measure whether the estimator is overfitted ($g > 1$) or underfitted ($g < 1$). However, since the aim of our exploration algorithm is to find a representative subset of the parameter space in the first place, we can instead consider $g$ as an estimated quality measure for our training set itself. For $g > 1$ the local prediction is better than the global prediction, hence the explored data can be seen as an oversimplified subset. For $g < 1$, conversely, the local prediction is worse than the global prediction and the explored data can be seen as overcomplicated subset. If $g = 1$, the local and global predictions are equally good. In this case the explored data set can be considered as perfectly representative.\par
The kernel SVM we use for classification allows us to almost always achieve a perfect score $\sigma$ and therefore we remain in the realm $g \geq 1$. Practice has shown that it seems to be a good approach to use such oversimplified data subsets during exploration. This observation could be due to the fact that our utility function is in such cases mostly based on the main features of the classification border and tends to neglect minor details, which are usually not important for all but the very last exploration steps.

\subsubsection{Ratios of false positives and false negatives}
Two additional global quality measures for our estimator are given by the ratio of false positives
\begin{subequations} \label{eqn:r}
\begin{align} \label{eqn:r:fp}
r_{\mathrm{fp}} \equiv r_{\mathrm{fp}}(D_{\mathrm{expl}})
\end{align}
and the ratio of false negatives
\begin{align} \label{eqn:r:fn}
r_{\mathrm{fn}} \equiv r_{\mathrm{fn}}(D_{\mathrm{expl}})
\end{align}
\end{subequations}
defined in~\cref{app:ratios of false positives and false negatives}. We use the convention that positive results correspond to valid outcomes and negative results to invalid outcomes. Consequently, $r_{\mathrm{fp}}$ represents the fraction of wrongly predicted valid outcomes and $r_{\mathrm{fn}}$ the fraction of wrongly predicted invalid outcomes, respectively, performed for almost all parameters in the parameter space $\boldsymbol{\chi}$.\par
An estimator of high quality is indicated by a high relative success rate $R$, \cref{eqn:R}, a low ratio of false positives $r_{\mathrm{fp}}$ and a low ratio of false negatives $r_{\mathrm{fn}}$. Depending on the application, either false positives or false negatives might be considered far more adverse. However, if the two ratios are to be valuated equally, considering $R$ might be a sufficient quality measure since by definition
\begin{align} \label{eqn:rR}
r_{\mathrm{fp}} + r_{\mathrm{fn}} = 1 - R
\end{align}
holds true.

\subsubsection{Validity ratio}
The calculation of valid data points yields meaningful optimization targets and therefore provides us with more information than the calculation of invalid data points. Hence, the fraction of valid to invalid data points can serve as a measure for the usefulness of a sampling. For this purpose we have defined the validity ratio
\begin{align} \label{eqn:alpha}
\alpha \equiv \alpha(D_{\mathrm{expl}})
\end{align}
as the fraction of valid to invalid data points in the explored data set $D_{\mathrm{expl}}$ in~\cref{app:validity ratio}. Moreover, we have defined its reference limit
\begin{align} \label{eqn:alphainf:toy}
\alpha_{\infty} = \frac{3\pi}{32 - 3\pi} \approx \num{0.417}
\end{align}
as the fraction of the total volumes for valid and invalid outcomes, respectively, in the whole parameter space. The validity ratio $\alpha$ of a uniform random sampling in the whole parameter space will eventually converge to $\alpha_{\infty}$ for a sufficient number of samples. Therefore, this value represents a reasonable reference scale for the validity rate $\alpha$ which other sampling approaches have to be measured up to.\par
The calculation of $\alpha_{\infty}$ also allows us to specify a worst-case reference limit for the ratios of false positives and false negatives, \cref{eqn:r}. Specifically, we consider a completely randomized estimator that predicts valid and invalid outcomes with equal chance. Performing such random predictions for almost all parameters in $\boldsymbol{\chi}$ lead us to the reference limit for the ratio of false positives
\begin{subequations} \label{eqn:rinf:toy}
\begin{align} \label{eqn:rinf:toy:p}
r_{\mathrm{fp} \infty} = \frac{3\pi}{64} \approx \num{0.147}
\end{align}
and the reference limit for the ratio of false negatives
\begin{align} \label{eqn:rinf:toy:n}
r_{\mathrm{fn} \infty} =  \frac{1}{2} - \frac{3\pi}{64} \approx \num{0.353},
\end{align}
\end{subequations}
respectively as explained in~\cref{app:validity ratio}. An estimator exceeding these limits is consequently worse than a ``random coin-tossing'' estimator.

\subsubsection{Exploration characteristics}
We list the characteristic values of exploration $R$, \cref{eqn:R}, $\sigma$, \cref{eqn:sigma}, $r_{\mathrm{fp}}$, \cref{eqn:r:fp}, $r_{\mathrm{fn}}$, \cref{eqn:r:fn}, and $\alpha$, \cref{eqn:alpha} together with the best and worst optimization targets
\begin{subequations} \label{eqn:tlimits}
\begin{align} \label{eqn:tlimits:best}
t_{\mathrm{best}} \equiv t_{\mathrm{max}}(D_{\mathrm{expl}})
\end{align}
and
\begin{align} \label{eqn:tlimits:worst}
t_{\mathrm{worst}} \equiv t_{\mathrm{min}}(D_{\mathrm{expl}}),
\end{align}
\end{subequations}
respectively, for each of the explored data sets, (1) to (5), in \cref{tab:toy-results}. In \Cref{eqn:tlimits} we have recalled the extremal optimization targets from \cref{eqn:textremal}.\par
A Monte Carlo approach is used to calculate
\begin{align} \label{eqn:R:approx}
R \approx R^{\mathrm{MC}}(D_{\mathrm{expl}}) \pm \delta R^{\mathrm{MC}}(D_{\mathrm{expl}})
\end{align}
with $\mathcal{N}=\num{10000}$ data points; see~\cref{app:relative success rate and score}. It is important to emphasize that global quality measures like $R$ are only possible because our considerations are not limited to predefined data sets, but rather make use of the fact that we can calculate new simulation data on demand. This enables us to create the randomly chosen samples necessary for the Monte Carlo approach.

\subsubsection{Discussion}
A comparison of the setups (2) and (5) in \cref{tab:toy-results} shows that our data exploration approach increases the highest relative success rate $R$ by almost $8\%$ in comparison with a regular grid sampling with the same number of sampling points. As expected, the highest value for $R$ is given by setup (3), followed by the setups (2) and (1). The worst performance is given by setup (4) and the grid approach, setup (5). However, while setup (4) has a poor value of $R$, it leads to the best optimization target $t_{\mathrm{best}} \approx \num{1.716}$, which almost reaches the theoretical limit of $\sqrt{3} \approx \num{1.732}$. This is no surprise given our choice of the weight vector $\mathbf{w}$, \cref{eqn:w}, which enforces parameter sampling near the best optimization target. Summarized, we see from our toy example that by tuning the weight vector we can intuitively control the exploration behavior.\par
The score $\sigma$ is perfect for all setups, which can be expected from such small training sets. Thus, we remain in the realm $g > 1$, \cref{eqn:g}.\par
Only the setups (2) and (3) have ratios $r_{\mathrm{fp}}$ below the worst-case reference limit $r_{\mathrm{fp} \infty}$, \cref{eqn:rinf:toy:p}, while the ratios of all other setups exceed it. The best ratio is achieved for setup (3) and the worst for setup (4), which incorporates the optimization target. The ratios $r_{\mathrm{fn}}$, on the other hand, are almost the same for all setups and are all much smaller than the worst-case reference limit $r_{\mathrm{fn} \infty}$, \cref{eqn:rinf:toy:n}. We assume that this result is due to the fact that invalid outcomes are mostly found in the outer realm of the parameter space, a topological behavior that can be uncovered with almost any sampling method even with only a few samples because of the low dimensionality of the parameter space.\par
A comparison of the validity ratios $\alpha$ reveals that the best result is achieved for setup (4), followed by (3) and (1). Setups (2) and (5) even fail to beat the reference limit $\alpha_{\infty}$, \cref{eqn:alphainf:toy}. As we will see further below, these relatively bad validity ratios are a consequence of the very small number of samples considered here and will improve with ongoing exploration. The fact that setup (1) has a better validity ratio than setup (2) is also a result of the sparse sampling.\par
For an exploration approach with a weight $o=0$, for which the optimization target is ignored, a lower value of $t_{\mathrm{worst}}$ can be considered more favorable since it indicates a more complete parameter space exploration. On the other hand, if the optimization target is of importance by choosing a weight $o>0$, exploration of such uninteresting regions should rather be avoided and a higher value of $t_{\mathrm{worst}}$ can be considered more favorable. The latter is the case for setup (4) and we find that it shares the same value for $t_{\mathrm{worst}}$ with setup (1) and the grid approach, setup (5).

\subsubsection{Summary}
The toy example clearly shows how the weights in the utility function can be used to control the explorative behavior in an intuitive way. For an accomplished data exploration, we consider a high value of $R$, $\alpha$ and $t_{\mathrm{best}}$ as desirable results, whereas $r_{\mathrm{fp}}$ and $r_{\mathrm{fn}}$ should be small. In other words, we seek (i) a good outcome prediction while (ii) evaluations in regions with a good optimization score should be preferred and (iii) the explored data set has a good ratio of valid to invalid outcomes. In this sense, these five exploration characteristics can be seen as objectives of a MCO problem with possibly conflicting goals. Relative importance of these objectives varies by the application.

\begin{figure}
\centering\includegraphics{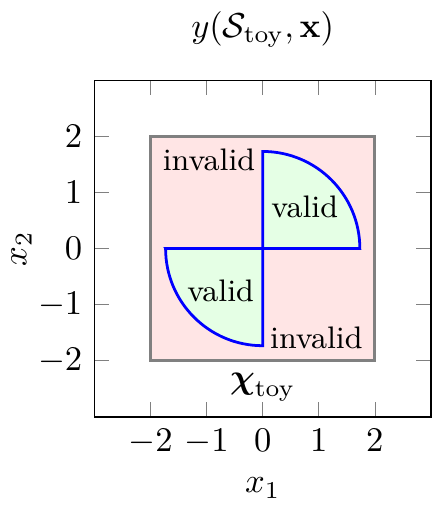}
\caption{Outcomes $y(\mathcal{S}_{\mathrm{toy}},\mathbf{x})$ of the toy simulation $\mathcal{S}_{\mathrm{toy}}$, \cref{eqn:toy}, in the toy example parameter space $\boldsymbol{\chi}_{\mathrm{toy}}$, \cref{eqn:toy:chi}, with the parameters $x_1$ and $x_2$, \cref{eqn:toy:x}. By definition, outcomes are valid within and on a circle of radius $\sqrt{3}$ in the first and third quadrant and invalid otherwise.}\label{fig:toy-pspace}
\end{figure}

\begin{figure}
\centering\includegraphics{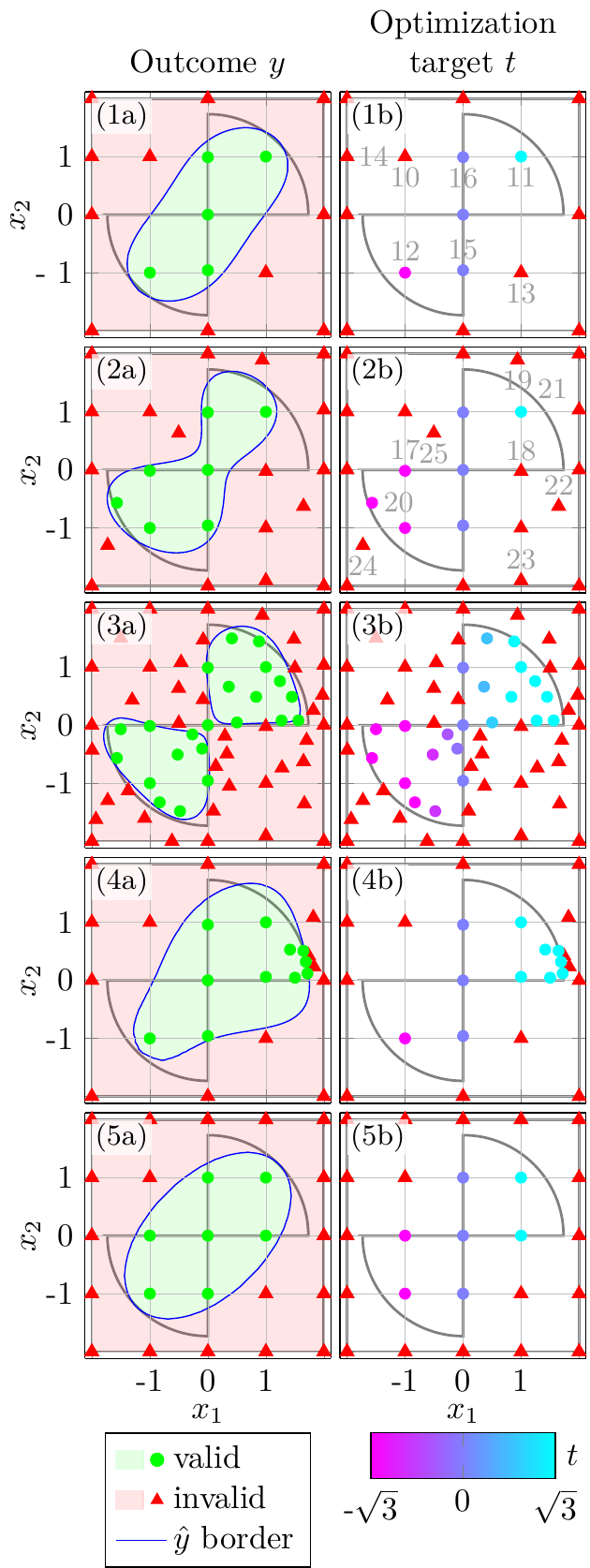}
\caption{Comparison of different data exploration setups for the toy example. Each row represents one of the exploration setups (1) to (5) from \cref{tab:toy-setups}. Column (a) shows the sampled outcomes $y$, \cref{eqn:toy:y}, with colored markers and the predicted outcomes $\hat{y}$, \cref{eqn:Emap}, with colored regions in the toy example parameter space $\boldsymbol{\chi}_{\mathrm{toy}}$, \cref{eqn:toy:chi}. Column (b) shows the respective targets $t$, \cref{eqn:toy:t}, for each sample. In both columns we also indicate the contours of the true feasibility region in analogy to \cref{fig:toy-pspace}. The numbers in (1b) and (2b) show the chronological order in which the points have been evaluated after the initial grid had been set up.}\label{fig:toy-explcomp}
\end{figure}

\begin{table}
\centering
\caption{Toy data exploration setups (1) to (5) with their respective exploration hyperparameters. We list the total number of samples $N$ and the parameter $G$ controlling the number of initial samples $G^2$, which are placed on a regular grid in $\boldsymbol{\chi}_{\mathrm{G}} = \boldsymbol{\chi}_{\mathrm{toy}}$, \cref{eqn:toy:chi}. The weights $s$, $o$ and $r$ constituting the weight vector $\mathbf{w}$, \cref{eqn:w}, determine the influence of the three components of the utility function, \cref{eqn:U}. Specifically, $s$ controls the outcome prediction uncertainty impact, $o$ the optimization target prediction impact and $r$ the distance to the nearest neighbor impact, respectively, \cref{eqn:u:interpretation}. Setups (1) to (4) use our algorithm, whereas setup (5) represents a pure grid approach in the framework of our algorithm, which is finished after the initial sampling step since $G^2 = N$.}\label{tab:toy-setups}
\begin{tabular}{|c||c|c|c|c|c|} \hline
\rule{0pt}{2.5ex} Setup & $N$ & $G$ & $s$ & $o$ & $r$ \\ \hline \hline
(1) & $16$ & $3$ & 1 & 0 & 1 \\ \hline
(2) & $25$ & $3$ & 1 & 0 & 1 \\ \hline
(3) & $64$ & $3$ & 1 & 0 & 1 \\ \hline
(4) & $25$ & $3$ & 1 & 3 & 1 \\ \hline
(5) & $25$ & $5$ & 0 & 0 & 0 \\ \hline
\end{tabular}
\end{table}

\begin{table}
\centering
\caption{Toy data exploration characteristics for each of the setups (1) to (5) from \cref{tab:toy-setups}. We list the success rate $R$, \cref{eqn:R:approx}, the score $\sigma$, \cref{eqn:sigma}, the ratio of false positives $r_{\mathrm{fp}}$, \cref{eqn:r:fp}, the ratio of false negatives $r_{\mathrm{fn}}$, \cref{eqn:r:fn}, the validity ratio $\alpha$, \cref{eqn:alpha}, the best optimization score $t_{\mathrm{best}}$, \cref{eqn:tlimits:best}, and the worst optimization score $t_{\mathrm{worst}}$, \cref{eqn:tlimits:worst}. The arrows indicate whether high values ($\uparrow$) or low values ($\downarrow$) are considered more favorable for each characteristic. Depending on the context, either a low value of $t_{\mathrm{worst}}$ (e.\,g., for $o=0$) or a high value (e.\,g., for $o>0$) can be desired.}\label{tab:toy-results}
\begin{tabular}{|c||c|c|c|c|c|c|c|} \hline
\rule{0pt}{2.5ex} Setup & $R$ $\uparrow$ & $\sigma$ $\uparrow$ & $r_{\mathrm{fp}}$ $\downarrow$ & $r_{\mathrm{fn}}$ $\downarrow$ & $\alpha$ $\uparrow$ & $t_{\mathrm{best}}$ $\uparrow$ & $t_{\mathrm{worst}}$ \\ \hline \hline
(1) & \num{0.807 +- 0.008} & \num{1.000} & \num{0.167 +- 0.007} & \num{0.026 +- 0.003} & \num{0.455} & \num{1.000} & \num{-1.000} \\ \hline
(2) & \num{0.857 +- 0.007} & \num{1.000} & \num{0.102 +- 0.004} & \num{0.041 +- 0.004} & \num{0.389}& \num{1.000} & \num{-1.568} \\ \hline
(3) & \num{0.952 +- 0.004} & \num{1.000} & \num{0.039 +- 0.004} & \num{0.009 +- 0.002} & \num{0.524}& \num{1.561} & \num{-1.568} \\ \hline
(4) & \num{0.761 +- 0.008} & \num{1.000} & \num{0.220 +- 0.007} & \num{0.019 +- 0.003} & \num{0.786}& \num{1.716} & \num{-1.000} \\ \hline
(5) & \num{0.797 +- 0.008} & \num{1.000} & \num{0.178 +- 0.007} & \num{0.025 +- 0.003} & \num{0.389}& \num{1.000} & \num{-1.000} \\ \hline
\end{tabular}
\end{table}

\section{Benchmark} \label{sec:Benchmark}
To demonstrate the strengths of our algorithm, we will in the following briefly present a benchmark with a Kriging-based exploration approach, which has been described in Refs.~\cite{Boukouvala2012,Wang2017}. We have already briefly mentioned this alternative strategy in \cref{sec:Introduction}. Summarized, the Kriging-based exploration works in an iterative way similar to our novel method: Starting from an initial data set, a feasibility estimator is trained each step with the currently explored data points. However, in contrast to our algorithm, the estimator is continuous and therefore relies on a continuous feasibility function which reflects the degree of feasibility constraint violation. Based on the continuous estimator, a new parameter is suggested and the simulation is evaluated for the new parameter. The resulting data point is included in the set of explored data points and the next iteration begins.\par
A more detailed explanation of the Kriging-based exploration can be found in Refs.~\cite{Boukouvala2012,Wang2017}. Our algorithm is described in \cref{sec:Optimized data exploration}. For the benchmark we will make use of our toy simulation, \cref{eqn:toy}, from \cref{sec:Demonstration}.

\subsection{Competing algorithms}
The main purpose of the benchmark is to study the behavior of the two algorithms of interest -- our proposed method and the Kriging-based approach -- with respect to the number of sampled points $N$. Therefore, we run our algorithm on a collection of setups $(6)_n$ defined by the hyperparameters $G = 4$, $\boldsymbol{\chi}_{\mathrm{G}} = \boldsymbol{\chi}_{\mathrm{toy}}$, $s=2$, $o=0$ and $r=1$. By setting $s > r$ we suppressed the spreading of data points in favor of a more precise feasibility border sampling. Each setup in the collection only differs by its number of samples $N=n$.\par
The notation of the setups follows \cref{sec:Demonstration}. Furthermore, we use an additional convention: The type of brackets we use to denote an exploration setup stands for the type of algorithm this setup is using to determine the sampling. Setups with round brackets refer to our algorithm, whereas non-round brackets represent different algorithms, as we will see in the following.

\subsubsection{Kriging-based approach}
For comparison, we run the Kriging-based exploration algorithm on a collection of setups $[7]_n$, where $n$ stands for the total number of sampled points in the same sense as for the setups $(6)_n$. For the estimator we use a Gaussian Process Regression (GPR)~\cite{Rasmussen2006} and achieved the best results using a Mat{\'e}rn kernel~\cite{Matern2013} with smoothness parameter $\nu=\num{1.5}$. In each iteration the hyperparameters of the GPR kernel are optimized by maximizing the log-marginal-likelihood of the GPR model with the help of the L-BFGS-B algorithm~\cite{Byrd1995}. The training data is standardized by removing the mean and scaling to unit variance. Because the toy simulation, \cref{eqn:toy}, is defined as a binary classification problem and by design no explicit information about feasibility constraint violation is available, we use
\begin{equation} \label{eqn:gpr:y}
y^{\mathrm{GPR}}(\mathbf{x}) \equiv \begin{cases} -1 & \text{if} \ y(\mathcal{S}_{\mathrm{toy}},\mathbf{x}) = \mathrm{valid} \\ 1 & \text{otherwise} \end{cases}
\end{equation}
as target values for the GPR model. In each iteration step, a new parameter $\textbf{x}^{\mathrm{GPR}}_{ \mathrm{new}} \in \boldsymbol{\chi}$ is sampled where the expected improvement~\cite{Wang2017}
\begin{equation} \label{eqn:gpr:EI}
\hat{I}^{\mathrm{GPR}}(\textbf{x}) \equiv \frac{s^{\mathrm{GPR}}(\textbf{x})}{\sqrt{2 \pi}} \exp\left(-\frac{\left[ \hat{y}^{\mathrm{GPR}}(\textbf{x}) \right]^2}{2 \left[ s^{\mathrm{GPR}}(\textbf{x}) \right]^2}\right)
\end{equation}
becomes maximal so that
\begin{equation} \label{eqn:gpr:xnew}
\textbf{x}^{\mathrm{GPR}}_{ \mathrm{new}} \equiv \argmax_{\textbf{x} \in \boldsymbol{\chi}} \hat{I}^{\mathrm{GPR}}(\textbf{x})
\end{equation}
holds true. Here we have made use of the GPR model prediction $\hat{y}^{\mathrm{GPR}}(\textbf{x})$ and its corresponding standard error $s^{\mathrm{GPR}}(\textbf{x})$ evaluated for the parameter  $\textbf{x}$. The initial data set for the training of the GPR model is chosen as a regular grid of $4 \times 4$ data points in complete analogy to the initial data set for our algorithm. The numerical solution of \cref{eqn:gpr:xnew} is obtained with a differential evolution approach.

\subsubsection{Modified Kriging-based approach}
The expected improvement, \cref{eqn:gpr:EI}, is supposed to focus the sampling on the feasibility border where the predicted violation $\hat{y}^{\mathrm{GPR}}(\textbf{x})$ vanishes. However, in our binary classification scenario with the constraint violation described by~\cref{eqn:gpr:y}, all points are penalized by either one of the two outcomes regardless of their distance to the border. This lack of knowledge about a continuous distance measure might significantly worsen the performance of the Kriging-based approach. Therefore, we suggest a modification which can compensate this deficiency. Specifically, we consider a reformulation
\begin{equation} \label{eqn:gpr:ycontinuous}
y^{\mathrm{GPR}} _{\mathrm{continuous}}(\mathbf{x}) \equiv \begin{cases} -\mathrm{d_{fb}}(\mathbf{x}) & \text{if} \ y(\mathcal{S}_{\mathrm{toy}},\mathbf{x}) = \mathrm{valid} \\ \mathrm{d_{fb}}(\mathbf{x})  & \text{otherwise} \end{cases}
\end{equation}
of~\cref{eqn:gpr:y}, where
\begin{equation} \label{eqn:gpr:ycontinuous:d}
\mathrm{d_{fb}}(\mathbf{x}) \equiv \underset{\substack{\mathbf{x}^{\prime} \in D_{\mathrm{expl}} \\ \text{s.\,t.} \ \ y(\mathcal{S}_{\mathrm{toy}},\mathbf{x}^{\prime}) \neq y(\mathcal{S}_{\mathrm{toy}},\mathbf{x})}}{\min} ||\mathbf{x} - \mathbf{x}^{\prime}||_2
\end{equation}
represents an upper bound of the closest Euclidean distance of the parameter $\mathbf{x}$ to the feasibility border. According to \cref{eqn:gpr:ycontinuous:d}, this upper bound corresponds to the Euclidean distance of $\mathbf{x}$ to its nearest neighbor of opposing feasibility $\mathbf{x}^{\prime}$ in the currently explored data set $D_{\mathrm{expl}}$. In particular, additional samples in $D_{\mathrm{expl}}$ can only improve $\mathrm{d_{fb}}(\mathbf{x})$, which converges to a tight bound for the theoretical limit of an infinite number of samples.\par
Apart from the major difference of using $y^{\mathrm{GPR}}_{\mathrm{continuous}}(\mathbf{x})$ instead of $y^{\mathrm{GPR}}(\mathbf{x})$ we leave the Kriging-based algorithm unchanged. For comparison, we run this modified algorithm on a collection of setups $\{8\}_n$, where $n$ stands for the total number of sampled points in the same sense as for the setups $(6)_n$ and $[7]_n$.

\subsection{Results}
For the benchmark we compare the exploration characteristics introduced in~\cref{sec:Demonstration:Results}. Specifically, we consider the success rate $R$, \cref{eqn:R:approx}, the score $\sigma$, \cref{eqn:sigma}, the ratio of false positives $r_{\mathrm{fp}}$, \cref{eqn:r:fp}, the ratio of false negatives $r_{\mathrm{fn}}$, \cref{eqn:r:fn}, and the validity ratio $\alpha$, \cref{eqn:alpha}, together with the best and worst optimization scores $t_{\mathrm{best}}$ and $t_{\mathrm{worst}}$, respectively, \cref{eqn:tlimits}. The exploration characteristics are summarized in~\cref{tab:bm-results} for four different numbers of samples $n \in \{ 25, 50, 100, 150 \}$ for each of the three candidate algorithms. Additionally, we show the progression of certain characteristics for an increasing number of samples $n \in [20, 150]$ in~\cref{fig:bm-results-progression}. The vertical lines (\plotline{gray,dotted}) correspond to the four chosen numbers of samples from~\cref{tab:bm-results}. Finally, snapshots of the explored outcomes for these samples can be found in~\cref{fig:bm-sample} analogously to the left column in~\cref{fig:toy-explcomp}.

\subsubsection{Discussion}
\begin{figure}
\centering\includegraphics{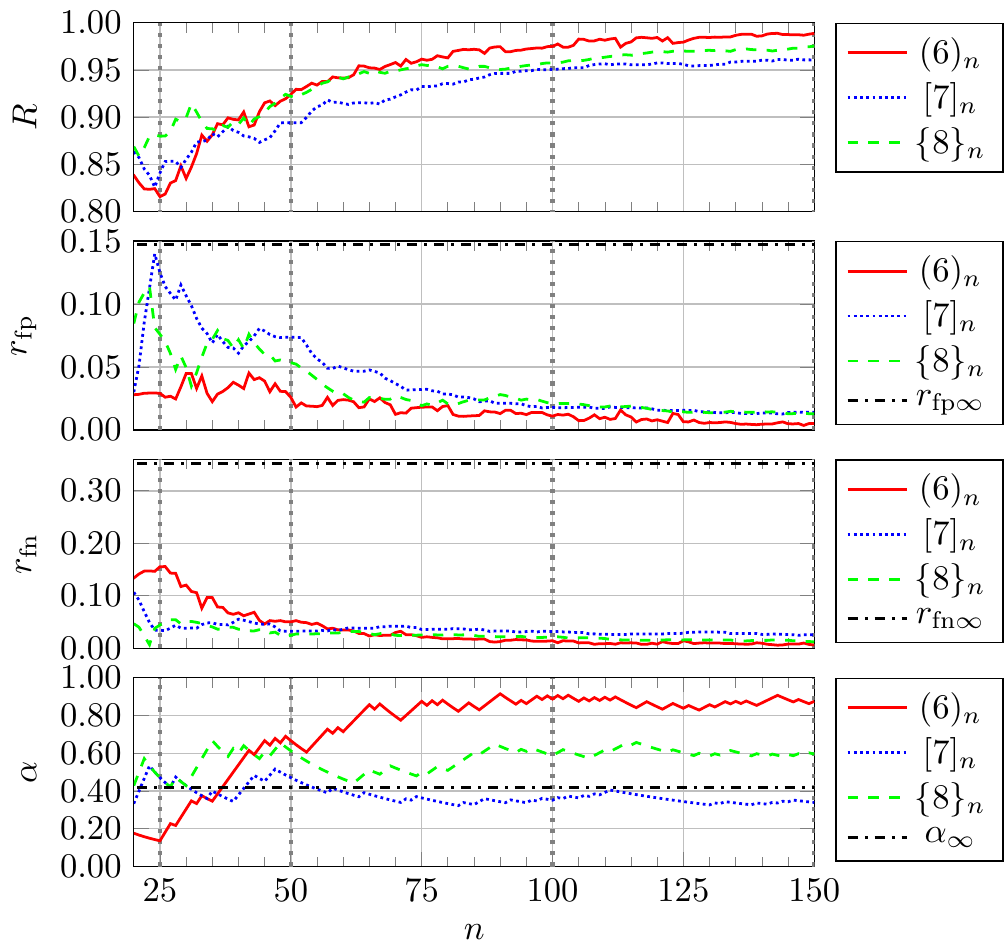}
\caption{Progression of benchmark exploration characteristics and their reference limits for a different number of samples $n$. We show the success rate $R$, \cref{eqn:R:approx}, the ratio of false positives $r_{\mathrm{fp}}$, \cref{eqn:r:fp}, the ratio of false negatives $r_{\mathrm{fn}}$, \cref{eqn:r:fn}, and the validity ratio $\alpha$. Each plot shows the results for the setup collections $(6)_n$, which use our algorithm, the setup collection $[7]_n$, which uses the Kriging-based algorithm, and the setup collection $\{8\}_n$, which uses the modified Kriging algorithm. The vertical lines mark the four chosen numbers of samples $n \in \{ 25, 50, 100, 150 \}$ shown in~\cref{tab:bm-results,fig:bm-sample}. Our algorithm clearly outscores the Kriging-based approach for a sufficient number of samples, in between lies the performance of the modified Kriging-based approach.}\label{fig:bm-results-progression}
\end{figure}

\begin{figure}
\centering\includegraphics{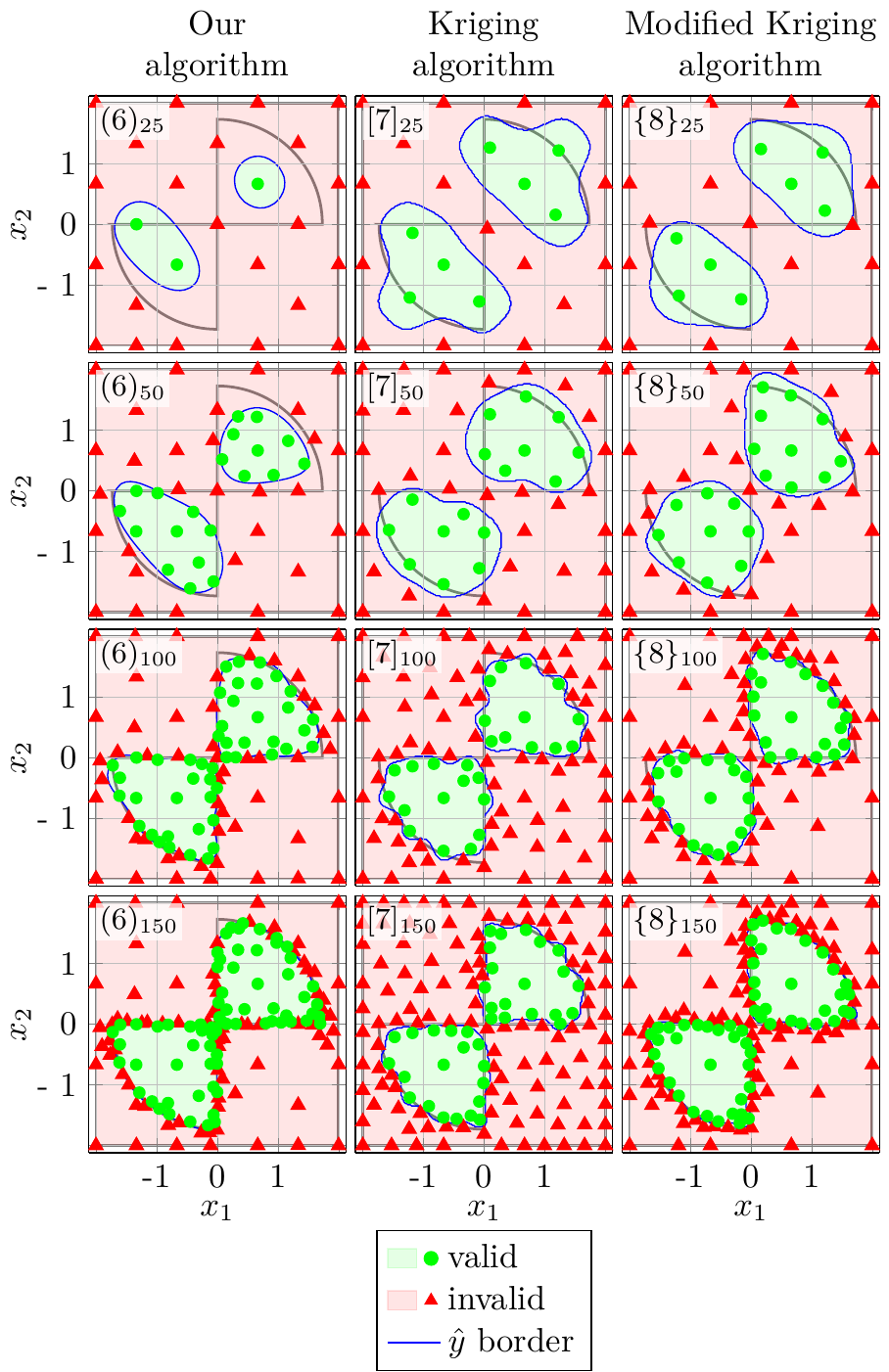}
\caption{Comparison of different data exploration setups for the benchmark in analogy to column (a) in~\cref{fig:toy-explcomp}. Each of the eight plots shows the sampled outcomes $y$, \cref{eqn:toy:y}, and the predicted outcomes $\hat{y}$, \cref{eqn:Emap}, in the toy example parameter space $\boldsymbol{\chi}_{\mathrm{toy}}$, \cref{eqn:toy:chi}, for a single setup. The left column shows the setups $(6)_{25}$, $(6)_{50}$, $(6)_{100}$ and $(6)_{150}$, which use our algorithm. The middle column shows the setups $[7]_{25}$, $[7]_{50}$, $[7]_{100}$ and $[7]_{150}$, which use the Kriging-based algorithm. The right column shows the setups $\{8\}_{25}$, $\{8\}_{50}$, $\{8\}_{100}$ and $\{8\}_{150}$, which use the modified Kriging-based algorithm. For each plot we indicate the contours of the true feasibility region in analogy to~\cref{fig:toy-pspace}.}\label{fig:bm-sample}
\end{figure}

The top plot in~\cref{fig:bm-results-progression} shows that after about 35 samples our algorithm (\plotline{red,solid}) steadily exceeds the success rate $R$ of the Kriging-based method (\plotline{blue,dotted}) by roughly $\num{0.02}$, and similarly, by roughly $0.01$ the success rate $R$ of the modified Kriging algorithm (\plotline{green,dashed}), albeit after about 80 samples. Both Kriging models exhibit higher $R$ values initially, however at the cost of increased false positive rates $r_{\mathrm{fp}}$.
The false positive rate is consistently better for our algorithm over all number of samples. Both Kriging-based algorithms exhibit a peak in the false positive rate in the initial sampling stage up, even up to the reference limit $r_{\mathrm{fp} \infty}$, \cref{eqn:rinf:toy:p}, indicating overestimation of the feasible range as illustrated in the top row of~\cref{fig:bm-sample}. On the other hand, the third plot in~\cref{fig:bm-results-progression} shows that our algorithm tends to underestimate the feasible region, especially for the first 35 samples. However, the major difference between benchmarked algorithms is highlighted in the bottom plot in~\cref{fig:bm-results-progression}. Our algorithm exceeds the reference validity ratio $\alpha_{\infty}$, \cref{eqn:alphainf:toy} already after about 35 samples. After about 70 samples the validity ratio of our algorithm is twice as high as that of the Kriging-based algorithm, and about $4/3$ times higher than that of the modified Kriging algorithm. Such high validity ratio values indicate that majority of new points are placed within feasible region and thus relevant parameter space is sampled more efficiently. Without modification, the Kriging-based algorithm reaches the validity ratio of a uniform sampling, thus confirming that the binary constraint violation formulation in~\cref{eqn:gpr:y} does not penalize sufficiently sampling far from the feasibility border. The modified constraint violation function, \cref{eqn:gpr:ycontinuous}, appears to partially address this issue as it results in an increased validity ratio, albeit it does not reach the level of our algorithm. \par
Our algorithm constantly pushes $t_{\mathrm{best}}$ and $t_{\mathrm{worst}}$ to their limits with an increasing number of samples, whereas the Kriging-based algorithm already reaches fixed values after about 25 sampled points. Those fixed values are far from the theoretical limits given by $\pm\sqrt{3} \approx \pm\num{1.732}$. This indicates that an exploration of the feasibility region is rather coarse with the Kriging-based approach. The modified Kriging algorithm reaches similar  $t_{\mathrm{best}}$ and $t_{\mathrm{worst}}$ values indicating better sampling at the feasible boundary than the unmodified Kriging-based algorithm. Note that we have not included the optimization target in our utility function, which would lead to a better value of $t_{\mathrm{best}}$ for much less samples as shown in \cref{sec:Demonstration:Results}. Moreover, the optimization target is not explicitly considered in the Kriging-based algorithms in the first place. The best optimization target can therefore be expected to be much worse if the target maximum is not directly located on the feasibility border.\par
From the progression of $R$ in~\cref{fig:bm-results-progression} we see that the prediction quality of all compared algorithms appears to be almost saturated for 150 samples. Although additional data points can still lead to an improvement of the estimators' precision, this improvement is expected to be rather small and might not be worth the additional calculation effort. According to~\cref{tab:bm-results} we have $\sigma < R$ so that $g < 1$, \cref{eqn:g}, for $(6)_{150}$, which indicates that the local prediction is worse than the global prediction. Thus, the explored data set can be seen as an overcomplicated subset of the complete parameter space. Such a behavior could in fact be used as a possible quantitative stopping criterion for our algorithm in agreement with the qualitatively observed saturation.\par

\subsubsection{Summary}
The benchmark has shown that our method is comparable to the Kriging-based approach for a very small number of samples, but outperforms it for an increasing number of iterations. It is particularly remarkable that we achieve a much higher ratio of valid to invalid data points with our algorithm, which makes our sampling more efficient. Moreover, we can clearly consider our modification of the Kriging-based approach as an improvement for the case of a discrete feasibility constraint violation.

\begin{table}
\centering
\caption{Benchmark exploration characteristics for the three setup collections $(6)_n$, $[7]_n$, and $\{8\}_n$, where $n$ represents the total number of sampled points $N$. The setups $(6)_{25}$, $(6)_{50}$, $(6)_{100}$ and $(6)_{150}$ use our algorithm, whereas $[7]_{25}$, $[7]_{50}$, $[7]_{100}$ and $[7]_{150}$ use the Kriging-based algorithm. The setups $\{8\}_{25}$, $\{8\}_{50}$, $\{8\}_{100}$ and $\{8\}_{150}$ use the modified Kriging-based algorithm. For each setup, we show the success rate $R$, \cref{eqn:R:approx}, the score $\sigma$, \cref{eqn:sigma}, the ratio of false positives $r_{\mathrm{fp}}$, \cref{eqn:r:fp}, the ratio of false negatives $r_{\mathrm{fn}}$, \cref{eqn:r:fn}, the validity ratio $\alpha$, \cref{eqn:alpha}, the best optimization score $t_{\mathrm{best}}$, \cref{eqn:tlimits:best}, and the worst optimization score $t_{\mathrm{worst}}$, \cref{eqn:tlimits:worst}. The arrows indicate whether high values ($\uparrow$) or low values ($\downarrow$) are considered more favorable for each characteristic. Depending on the context, either a low or a high value of $t_{\mathrm{worst}}$ might be preferable. According to these characteristics, our algorithm is comparable to the Kriging-based approach for few samples, but superior for many. The modified Kriging-based approach generally performs better than the original version. A visualization of exploration characteristics for different sample sizes can be found in \cref{fig:bm-results-progression}.}\label{tab:bm-results}
\begin{tabular}{|c||c|c|c|c|c|c|c|c|c|} \hline
\rule{0pt}{2.5ex} Setup & $R$ $\uparrow$ & $\sigma$ $\uparrow$ & $r_{\mathrm{fp}}$ $\downarrow$ & $r_{\mathrm{fn}}$ $\downarrow$ & $\alpha$ $\uparrow$ & $t_{\mathrm{best}}$ $\uparrow$ & $t_{\mathrm{worst}}$ \\ \hline \hline
$(6)_{25}$ & \num{0.816 +- 0.008} & \num{1.000} & \num{0.029 +- 0.003} & \num{0.155 +- 0.007} & \num{0.136}  & \num{0.667} & \num{-1.333} \\ \hline
$(6)_{50}$ & \num{0.924 +- 0.005} & \num{1.000} & \num{0.026 +- 0.003} & \num{0.050 +- 0.004} & \num{0.667}  & \num{1.435} & \num{-1.606} \\ \hline
$(6)_{100}$ & \num{0.975 +- 0.003} & \num{1.000} & \num{0.011 +- 0.002} & \num{0.014 +- 0.002} & \num{0.887}  & \num{1.580} & \num{-1.723} \\ \hline
$(6)_{150}$ & \num{0.988 +- 0.002} & \num{0.973} & \num{0.005 +- 0.001} & \num{0.006 +- 0.002} & \num{0.875}  & \num{1.696} & \num{-1.724} \\ \hline
$[7]_{25}$ & \num{0.841 +- 0.007} & \num{1.000} & \num{0.125 +- 0.006} & \num{0.034 +- 0.004} & \num{0.471}  & \num{1.229} & \num{-1.225} \\ \hline
$[7]_{50}$ & \num{0.894 +- 0.006} & \num{1.000} & \num{0.074 +- 0.005} & \num{0.032 +- 0.003} & \num{0.471}  & \num{1.558} & \num{-1.568} \\ \hline
$[7]_{100}$ & \num{0.950 +- 0.004} & \num{1.000} & \num{0.018 +- 0.003} & \num{0.032 +- 0.003} & \num{0.351} & \num{1.558} & \num{-1.568} \\ \hline
$[7]_{150}$ & \num{0.960 +- 0.004} & \num{1.000} & \num{0.014 +- 0.002} & \num{0.026 +- 0.003} & \num{0.339} & \num{1.558} & \num{-1.568}\\ \hline
$\{8\}_{25}$ & \num{0.880 +- 0.006} & \num{1.000} & \num{0.076 +- 0.005} & \num{0.044 +- 0.004} & \num{0.471}  & \num{1.220} & \num{-1.226} \\ \hline
$\{8\}_{50}$ & \num{0.922 +- 0.005} & \num{1.000} & \num{0.053 +- 0.004} & \num{0.025 +- 0.003} & \num{0.613}  & \num{1.481} & \num{-1.524} \\ \hline
$\{8\}_{100}$ & \num{0.957 +- 0.004} & \num{1.000} & \num{0.021 +- 0.003} & \num{0.022 +- 0.003} & \num{0.587}  & \num{1.571} & \num{-1.530} \\ \hline
$\{8\}_{150}$ & \num{0.975 +- 0.003} & \num{0.993} & \num{0.013 +- 0.002} & \num{0.012 +- 0.002} & \num{0.596}  & \num{1.631} & \num{-1.645} \\ \hline
\end{tabular}
\end{table}

\section{Application} \label{sec:Application}
In the current section we will apply the proposed algorithm from~\cref{sec:Optimized data exploration} to the simulation of a realistic chemical process. This process describes the inner workings of a production plant consisting of two connected distillation columns. The task of the plant is to separate a two-component mixture consisting of chloroform and acetone by means of a so-called pressure swing distillation. Such kind of operations are very common in chemical engineering and their physical description is well-known~\cite{Biegler1997}.\par
As shown in~\cref{fig:plant} each column has one input stream (feed) and two output streams (distillate and bottom stream), consisting of chloroform and acetone mixtures. Different parameters such as the operating pressure $P$ influence the concentrations of the components in the output streams. The two columns are connected in such a way that the bottom stream of the first column constitutes the feed for the second column and the bottom stream of the second column is recycled and mixed with the educt of the process to form the feed of the first column. The two distillate streams constitute the products of the process, which are desired to have high purities.\par
The distillation process we consider is not straightforward since the mixture of chloroform and acetone exhibits an azeotropic behavior. This means that at the azeotropic point, the composition in the vapor phase equals the composition in the liquid phase, which is sensitive to $P$ in this particular case. One possibility to surpass this distillation limit and to separate the azeotropic mixture is to use a special setup in which the first column operates at low pressure ($P=\SI{1}{\bar}$) and the second column operates at high pressure ($P=\SI{10}{\bar}$). As a result, a high concentration of acetone in the distillate stream of the first column can be achieved and the concentrations in the corresponding bottom stream are close to the azeotrope. The higher pressure in the second column changes the azeotrope composition in such a way that a high concentration of chloroform can be achieved in the distillate stream of the second column. 

\subsection{Chemical simulation}
The most common approach for modeling distillation columns is the equilibrium stage model (ESM)~\cite{Biegler1997}, which is based on a cascade of interconnected equilibrium stages. Every stage has a vapor (boiling) and a liquid (condensing) output stream, where the vapor stream of each stage raises to the stage above and the liquid output stream of each stage flows to the stage below. The reflux ratio of a column represents the ratio between the internal liquid streams of the last two stages and the distillation stream. A higher reflux ratio means that more energy is needed for cooling and heating. For every stage certain physically motivated equations (conservation of mass and conservation of enthalpy together with thermodynamic equilibrium and closure conditions) must hold. The ESM is used in all commercially available flow sheet simulators and is used wold-wide to simulate chemical distillation processes.\par
\begin{figure}
\centering\includegraphics{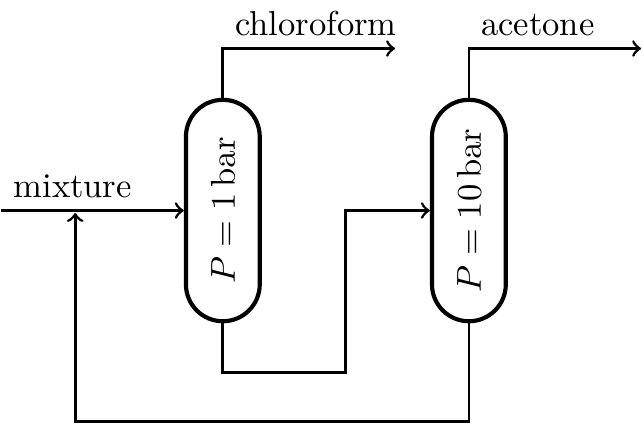}
\caption{Flowchart of the simulated swing distillation process consisting of two connected distillation columns with different pressures $P$. A mixture of chloroform and acetone enters the chemical plant as a continuous stream. The outgoing product consists of two separate distillate streams, one consisting of mostly chloroform and the other of mostly aceton. The highest possible purity of both streams is desired.}\label{fig:plant}
\end{figure}

We model each column with an ESM consisting of $\num{28}$ stages (not shown in~\cref{fig:plant}). The non-random two-liquid (NRTL) model~\cite{Renon1968} is used to describe the interactions between the substances in each stage. The complete chemical process is then represented by a system of about $\num{400}$ coupled linearly independent equations, some of which are highly nonlinear. We also have a comparable number of independent internal variables describing, e.\,g., the concentrations, temperatures and flow rates of the internal streams between the stages.\par
As already mentioned in~\cref{sec:Introduction}, we use the flowsheet simulator Chemasim to perform the calculations. Given the parameters $\mathbf{x}$, we formally define the function
\begin{align} \label{eqn:chem:y}
C(\mathbf{x}) & \equiv \begin{cases} \mathrm{valid} & \parbox[t]{.6\columnwidth}{if the Chemasim evaluation using $\mathbf{x}$ \\[-.2em] is well-defined and leads to a convergent \\[-.2em] and physically feasible result} \\ \mathrm{invalid} & \text{otherwise} \end{cases}
\end{align}
to describe the outcome of a Chemasim evaluation in the classification space $\eta$, \cref{eqn:eta}. Convergence and physical feasibility are exclusively decided by Chemasim-internal criteria and we have no knowledge about the degree of a violation, i.\,e., we have no access to a continuous feasibility function.\par
We consider a four-dimensional parameter space (i.\,e., $p=4$) for the chemical process simulation. It is constituted by the mass fractions of acetone at the distillate stream of column one $m_{\mathrm{ac}} \in [0.1,1.0]$ and chloroform at the distillate stream of column two $m_{\mathrm{cl}} \in [0.8,1.0]$, as well as the reflux ratios of the two columns $r_1 \in [5,35]$ and $r_2 \in [5,35]$, respectively. Using the parameter vector 
\begin{align} \label{eqn:chem:x}
\mathbf{x} \equiv \begin{pmatrix} m_{\mathrm{ac}} \\ m_{\mathrm{cl}} \\ r_1 \\ r_2 \end{pmatrix}
\end{align}
our parameter space can consequently be written as
\begin{align} \label{eqn:chem:chisim}
\boldsymbol{\chi}_{\mathrm{sim}} \equiv [0.1,1.0] \otimes [0.8,1.0] \otimes [5,35] \otimes [5,35].
\end{align}
and therefore corresponds to a four dimensional hyperrectangle.\par
The simulation $\mathcal{S}_{\mathrm{sim}}$, \cref{eqn:Smap}, can formally be expressed by
\begin{subequations} \label{eqn:chem}
\begin{align}
y(\mathcal{S}_{\mathrm{sim}},\mathbf{x}) \equiv C(\mathbf{x})
\end{align}
and
\begin{align} \label{eqn:chem:t}
t(\mathcal{S}_{\mathrm{sim}},\mathbf{x}) \equiv \frac{m_{\mathrm{ac}} + m_{\mathrm{cl}}}{2},
\end{align}
\end{subequations}
respectively, where we have recalled~\cref{eqn:chem:y}. Our optimization target $t(\mathcal{S}_{\mathrm{sim}},\mathbf{x})$ is chosen as the average of the two distillate mass fractions, which we aim to maximize. Recall that by definition, the optimization target is only meaningful for valid simulation outcomes $y(\mathcal{S}_{\mathrm{sim}},\mathbf{x})$. We specify sufficient internal variables of the simulation with a suitably chosen but fixed value so that given the parameters $\mathbf{x}$ the system of equations is well-defined and can be solved by Chemasim.\par
We assume that we have no previous knowledge about the data (i.\,e., $D_{\mathrm{init}} = \{\}$) except for the fact that there is a parameter space window
\begin{align} \label{eqn:chem:chiwin}
\boldsymbol{\chi}_{\mathrm{win}} \equiv [0.3,0.5] \otimes [0.875,0.925] \otimes [8,10] \otimes [8,10] \subset \boldsymbol{\chi}_{\mathrm{sim}}
\end{align}
in which both valid and invalid parameters can be found. For the start-up step of our algorithm we choose a regular grid which expands inside of a chosen hyperrectangle $\boldsymbol{\chi}_{\mathrm{G}}$. \Cref{eqn:xnew} is again solved numerically using a differential evolution approach.

\subsection{Results}
\begin{figure}
\centering\includegraphics{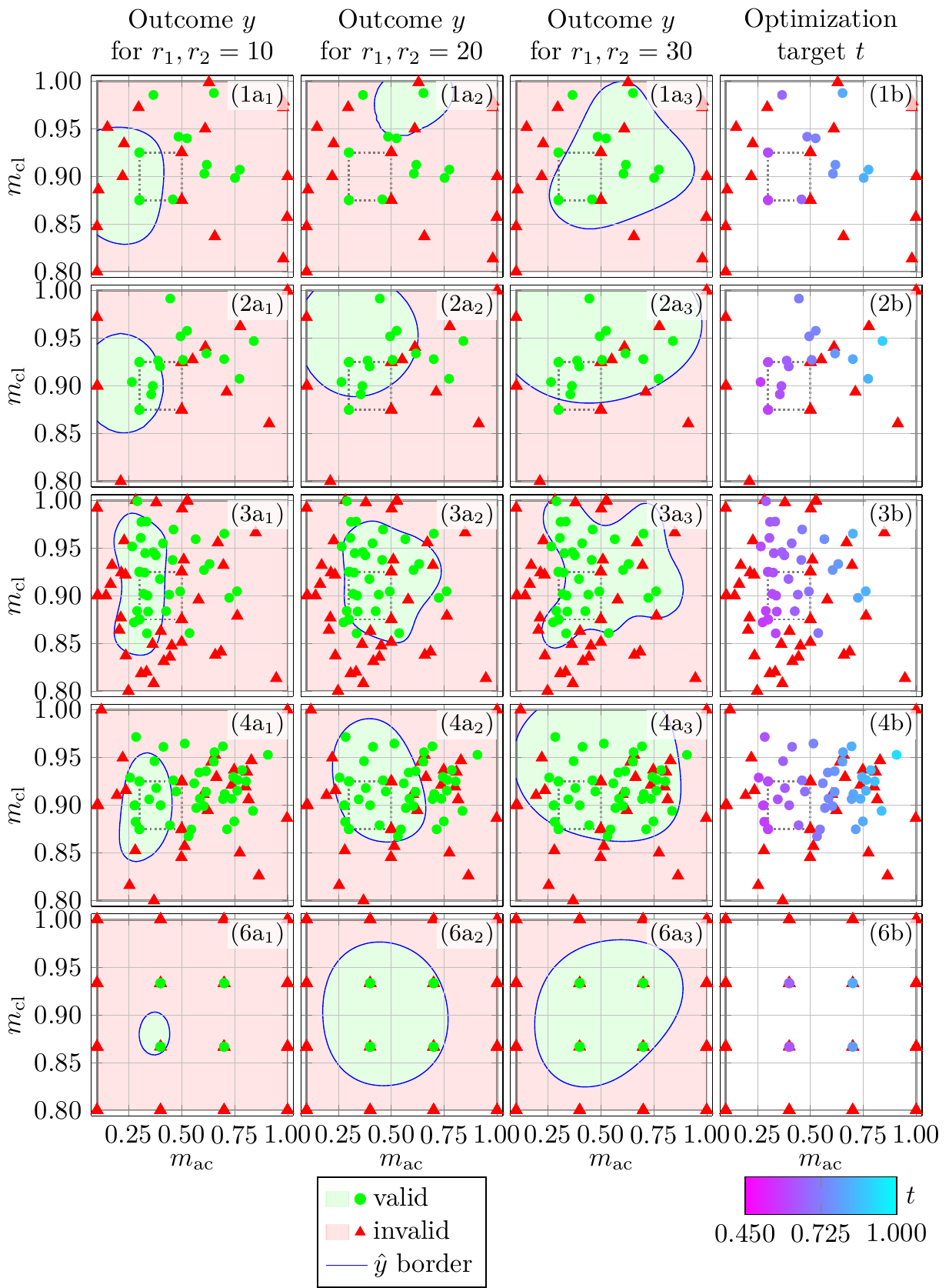}
\caption{Comparison of different data exploration setups for the chemical simulation in analogy to~\cref{fig:toy-explcomp}. In each of the first three columns we show a two dimensional planar cut through the four dimensional parameter space $\boldsymbol{\chi}_{\mathrm{sim}}$, \cref{eqn:chem:chisim}, and project all data points onto each plane. We choose the planes for which \cref{eqn:r12-plane} holds true with $c=10$ in column ($\mathrm{a_1}$), $c=20$ in column ($\mathrm{a_2}$) and $c=30$ in column ($\mathrm{a_3}$), respectively. The outcomes of the projected data points do not necessarily have to correspond to the estimated feasibility border on the plane. The exact theoretical feasibility border is unknown and cannot be plotted. The dotted rectangle shows the parameter space window $\boldsymbol{\chi}_{\mathrm{win}}$, \cref{eqn:chem:chiwin}, used for the initial grid sampling.}\label{fig:chem-explcomp}
\end{figure}

The results are shown in~\cref{fig:chem-explcomp} analogously to~\cref{fig:toy-explcomp}. To achieve a two dimensional illustration of the four dimensional parameter space $\boldsymbol{\chi}_{\mathrm{sim}}$, \cref{eqn:chem:chisim}, we show three different planar cuts through $\boldsymbol{\chi}_{\mathrm{sim}}$ in the first three columns and project all data points onto the respective planes. Specifically, we choose the planes for which
\begin{align} \label{eqn:r12-plane}
r_1=r_2=c
\end{align}
holds true with $c=10$ in column ($\mathrm{a_1}$), $c=20$ in column ($\mathrm{a_2}$) and $c=30$ in column ($\mathrm{a_3}$), respectively. As a consequence, the outcomes of the projected data points do not necessarily have to correspond to the estimated feasibility borders (\plotline{blue}) on each plane. It is also important to emphasize that the exact theoretical feasibility border is unknown so we cannot plot it. For the optimization target visualization in column (b), the specific choice of the projection plane does not affect the plots.\par
Each of the five rows represents a different setup as summarized in~\cref{tab:chem-setups}. Specifically, the first two setups are based on $N=40$ evaluations and while setup (1) ignores the optimization target, it is taken into account by setup (2) with an equal weight. The third and fourth setup also contain these two cases but for $N=81$ evaluations. All of these four setups have an initial grid parameter $G=2$ and make use of a grid in the parameter space window $\boldsymbol{\chi}_{\mathrm{win}}$, \cref{eqn:chem:chiwin}. Finally, the setups (5) and (6) represent pure grid approaches in the framework of our algorithm, which are finished after the initial sampling step since $G^4 = N$. In~\cref{tab:chem-setups} we also list three additional setups, $\langle$7$\rangle$ to $\langle$9$\rangle$, which are not shown in~\cref{fig:chem-explcomp}. Each of these setups represents a typical LHS of $N$ parameter points in $\boldsymbol{\chi}_{\mathrm{sim}}$, \cref{eqn:chem:chisim}. The other exploration hyperparameters have no meaning for these setups.\par
As one would expect, it becomes apparent from columns ($\mathrm{a_1}$) to ($\mathrm{a_3}$) that the estimated feasible regions expand and include higher distillate mass fractions $m_{\mathrm{ac}}$ and $m_{\mathrm{cl}}$, respectively, with increased reflux ratios $r_1=r_2$. Interestingly, this observation can be made for all depicted setups. The only exception can be found in ($1\mathrm{a_2}$), which covers a smaller validity area than ($1\mathrm{a_1}$), but still includes higher distillate mass fractions. Since $r_1,r_2 \in [5,35]$, column ($\mathrm{a_2}$) in fact shows a projection onto the central reflux plane.

\subsubsection{Exploration characteristics}
\Cref{fig:chem-explcomp} serves as an illustration of the exploration behavior, however, to quantify the success of individual setups, we compare the exploration characteristics introduced in~\cref{sec:Demonstration:Results}. The characteristic values of exploration are given by the success rate $R$, \cref{eqn:R}, the score $\sigma$, \cref{eqn:sigma}, the ratio of false positives $r_{\mathrm{fp}}$, \cref{eqn:r:fp}, the ratio of false negatives $r_{\mathrm{fn}}$, \cref{eqn:r:fn}, and the validity ratio $\alpha$, \cref{eqn:alpha}. We show these values together with the best and worst optimization scores $t_{\mathrm{best}}$ and $t_{\mathrm{worst}}$, respectively, \cref{eqn:tlimits}, in~\cref{tab:chem-results} for all nine setups.\par
For the calculation of $R$, $r_{\mathrm{fp}}$, $r_{\mathrm{fn}}$ and the respective reference limits $\alpha_{\infty}$, $r_{\mathrm{fp} \infty}$ and $r_{\mathrm{fn} \infty}$ we use a Monte Carlo approach with $\mathcal{N}=\num{10000}$ data points. Specifically, we make use of the success rate approximation, \cref{eqn:R:approx}, the approximation of the ratio of false positives
\begin{subequations} \label{eqn:r:approx}
\begin{align} \label{eqn:r:approx:fp}
r_{\mathrm{fp}} \approx r_{\mathrm{fp}}^{\mathrm{MC}}(D_{\mathrm{expl}}) \pm \delta r_{\mathrm{fp}}^{\mathrm{MC}}(D_{\mathrm{expl}})
\end{align}
and the approximation of the ratio of false negatives
\begin{align} \label{eqn:r:approx:fn}
r_{\mathrm{np}} \approx r_{\mathrm{np}}^{\mathrm{MC}}(D_{\mathrm{expl}}) \pm \delta r_{\mathrm{np}}^{\mathrm{MC}}(D_{\mathrm{expl}})
\end{align}
\end{subequations}
as explained in~\cref{app:relative success rate and score,app:ratios of false positives and false negatives}, respectively. Moreover, the above-mentioned reference limits are approximated by
\begin{align} \label{eqn:alphainf:approx}
\alpha_{\infty} \approx \alpha_{\infty}^{\mathrm{MC}} \pm \delta \alpha_{\infty}^{\mathrm{MC}},
\end{align}
\begin{subequations} \label{eqn:rinf:approx}
\begin{align}
r_{\mathrm{fp} \infty} \approx \frac{\alpha_{\infty}^{\mathrm{MC}}}{2(\alpha_{\infty}^{\mathrm{MC} } + 1)} \pm \delta r_{\infty}^{\mathrm{MC}}
\end{align}
and
\begin{align}
r_{\mathrm{fn} \infty} \approx \frac{1}{2(\alpha_{\infty}^{\mathrm{MC}} + 1)} \pm \delta r_{\infty}^{\mathrm{MC}}
\end{align}
\end{subequations}
as explained in~\cref{app:validity ratio,app:ratios of false positives and false negatives}, respectively. A numerical evaluation of~\cref{eqn:alphainf:approx} yields
\begin{align} \label{eqn:alphainf:approx:evaluated}
\alpha_{\infty} \approx \num{1.086 +- 0.043},
\end{align}
which indicates a close balance between valid and invalid points in the whole parameter space. Using~\cref{eqn:alphainf:approx:evaluated} we can directly determine
\begin{subequations} \label{eqn:rinf:approx:evaluated}
\begin{align} \label{eqn:rinf:approx:evaluated:p}
r_{\mathrm{fp} \infty} \approx \num{0.260 \pm 0.005}
\end{align}
and
\begin{align} \label{eqn:rinf:approx:evaluated:n}
r_{\mathrm{fn} \infty} \approx \num{0.240 \pm 0.005},
\end{align}
\end{subequations}
which are also very similar due to this balancing.

\subsubsection{Discussion}
We find that according to~\cref{tab:chem-results} setup (3) reaches the best value for the relative success rate $R$, closely followed by setup (1). Both results have overlapping confidence intervals. A slightly worse value for $R$ is achieved by the setups (2) and (4). By contrast, remarkably bad results can be found for the grid and LHS approaches. A comparison from setup (6) with setup (3) reveals that our data exploration approach results in an improvement of $R$ by almost $90\%$ with only one third of the sampling points ($81$ instead of $256$). This result shows that our algorithm is way more efficient especially for higher dimensional spaces than a uniform grid sampling.\par
The scores $\sigma$ show that the estimators trained on the LHS samples also perform bad on the training set itself. Since $\sigma$ increases with a higher number of samples, we assume that this behavior might be a consequence of the sparsity of the sample in the four dimensional parameter space, which can lead to an overly complex prediction of the feasibility border. It might therefore be possible that using a different estimator may result in a better overall performance for the LHS approaches. For all setups, we have $g > 1$, \cref{eqn:g}.\par
When we compare the ratios of false positives and false negatives for the first four setups, we find that setup (3) has the lowest value of $r_{\mathrm{fp}}$, followed by (1), (2) and (4). The lowest value of $r_{\mathrm{np}}$ is, on the other hand, achieved by (4), followed by (2), (1) and (3). It is remarkable that the setups (4), $\langle$8$\rangle$ and $\langle$9$\rangle$ exceed the reference limit $r_{\mathrm{fp} \infty}$, \cref{eqn:rinf:approx:evaluated:p}, but have almost no false negatives, whereas the setups (5), (6) and $\langle$7$\rangle$ exceed the reference limit $r_{\mathrm{np} \infty}$, \cref{eqn:rinf:approx:evaluated:n}, but have almost no false positives. Clearly, all of the associated estimators generalize very badly due to the unrepresentative sampling used for the training.\par
The validity ratio $\alpha$ is best for setup (4), followed by setup (2). Both of these ratios clearly exceed the reference limit $\alpha_{\infty}$, \cref{eqn:alphainf:approx:evaluated}. The other setups, including the three LHS approaches, only have validity ratios smaller than $\alpha_{\infty}$ and can therefore be considered less useful than a typical random sampling. This result is no surprise since respecting the optimization score will guide the exploration towards parameter space realms with valid outcomes.\par
As expected, we see from a comparison between (1) and (2) or (3) and (4) in~\cref{fig:chem-explcomp} that omission of the optimization target leads to a more regular spreading of the evaluations in the parameter space. Again, it is no surprise that incorporating the optimization score can lead to a worse outcome classification.\par
The highest value for $t_{\mathrm{best}}$ is reached by setup (4), closely followed by setup $\langle$9$\rangle$. All other setups achieve slightly worse values. Therefore, we find that even a small number of evaluations provides us with a reasonably good optimization target. From a direct optimization of the simulation by means of a MISQP procedure~\cite{Schittkowski2011} we find an optimization target of $t \approx 0.950$, which is only slightly better than our best exploration result of $t \approx 0.930$. Its calculation can, however, require a few hundred parameter evaluations. Moreover, the Chemasim-internal optimizer has difficulties solving this specific problem due to the fact that the optimization target, \cref{eqn:chem:t}, has vanishing derivatives with respect to $r_1$ and $r_2$, which makes them insignificant for a gradient-based optimization approach. The feasibility outcome, on the other hand, explicitly depends on those parameters. Since the optimizer ignores this relation, it can be very difficult to find an optimal target in Chemasim without injecting expert knowledge.\par
The value of $t_{\mathrm{worst}}$ shows us in how far regions of a rather uninteresting optimization score have been explored. Exploration of such regions might be necessary to improve $R$, but should be avoided in favor of more relevant regions if the optimization target is of interest. From a comparison between (3) and (4) we find, without surprise, that incorporating the optimization target into the utility function leads to a more favorable value of $t_{\mathrm{worst}}$. However, a comparison between (1) and (2) shows the opposite effect. We assume that this behavior is a result of an unfavorably trained estimator for the optimization target during one iteration of the exploration process.

\subsubsection{Summary}
Our data exploration method has proven to be clearly superior to grid-based or LHS approaches. From the calculation of comparably few data points we already get a very good estimation of the feasibility region in the whole parameter space. Moreover, by incorporating the optimization target in the utility function we can achieve a sampling which contains data points very close to the optimum. Such data points can serve as suitable starting points for a rigorous optimization algorithm.

\begin{table}
\centering
\caption{Data exploration setups for the chemical simulation. We list the setups (1) to (4), which use our algorithm, the setups (5) and (6), which represent a pure grid approach in the framework of our algorithm, and the setups $\langle$7$\rangle$ to $\langle$9$\rangle$, which are based on a LHS approach, with their respective exploration hyperparameters. For each setup, we show the total number of samples $N$. For setups (1) to (6) we also show and the parameter $G$ controlling the number of initial samples $G^4$, which are placed on a regular grid in $\boldsymbol{\chi}_{\mathrm{G}}$. The latter either corresponds to $\boldsymbol{\chi}_{\mathrm{sim}}$, \cref{eqn:chem:chisim}, or $\boldsymbol{\chi}_{\mathrm{win}}$, \cref{eqn:chem:chiwin}. The weights $s$, $o$ and $r$ determine the explorative behavior by changing the influence of the three components of the utility function, \cref{eqn:U}. Specifically, $s$ controls the outcome prediction uncertainty impact, $o$ the optimization target prediction impact and $r$ the distance to the nearest neighbor impact, respectively, \cref{eqn:u:interpretation}. All of these exploration hyperparameters have no meaning for the LHS setups.}\label{tab:chem-setups}
\begin{tabular}{|c||c|c|c|c|c|c|} \hline
\rule{0pt}{2.5ex} Setup & $N$ & $G$ & $\boldsymbol{\chi}_{\mathrm{G}}$ & $s$ & $o$ & $r$ \\ \hline \hline
(1) & $40$ & $2$ & $\boldsymbol{\chi}_{\mathrm{win}}$ & 1 & 0 & 1 \\ \hline
(2) & $40$ & $2$ & $\boldsymbol{\chi}_{\mathrm{win}}$ & 1 & 1 & 1 \\ \hline
(3) & $81$ & $2$ & $\boldsymbol{\chi}_{\mathrm{win}}$ & 1 & 0 & 1 \\ \hline
(4) & $81$ & $2$ & $\boldsymbol{\chi}_{\mathrm{win}}$ & 1 & 1 & 1 \\ \hline
(5) & $81$ & $3$ & $\boldsymbol{\chi}_{\mathrm{sim}}$ & 0 & 0 & 0 \\ \hline
(6) & $256$ & $4$ & $\boldsymbol{\chi}_{\mathrm{sim}}$ & 0 & 0 & 0 \\ \hline
$\langle$7$\rangle$ & $40$ & \multicolumn{5}{c|}{LHS in $\boldsymbol{\chi}_{\mathrm{sim}}$} \\ \hline
$\langle$8$\rangle$ & $81$ & \multicolumn{5}{c|}{LHS in $\boldsymbol{\chi}_{\mathrm{sim}}$} \\ \hline
$\langle$9$\rangle$ & $256$ & \multicolumn{5}{c|}{LHS in $\boldsymbol{\chi}_{\mathrm{sim}}$} \\ \hline
\end{tabular}
\end{table}

\begin{table}
\centering
\caption{Data exploration characteristics for the chemical simulation for each of the setups (1) to (4), which use our algorithm, the setups (5) and (6), which represent a pure grid approach in the framework of our algorithm, and the setups $\langle$7$\rangle$ to $\langle$9$\rangle$, which are based on a LHS approach. The respective hyperparameters are shown in~\cref{tab:chem-setups}. We list the success rate $R$, \cref{eqn:R:approx}, the score $\sigma$, \cref{eqn:sigma}, the ratio of false positives $r_{\mathrm{fp}}$, \cref{eqn:r:approx:fp}, the ratio of false negatives $r_{\mathrm{fn}}$, \cref{eqn:r:approx:fn}, the validity ratio $\alpha$, \cref{eqn:alpha}, the best optimization score $t_{\mathrm{best}}$, \cref{eqn:tlimits:best}, and the worst optimization score $t_{\mathrm{worst}}$, \cref{eqn:tlimits:worst}. The arrows indicate whether high values ($\uparrow$) or low values ($\downarrow$) are considered more favorable for each characteristic. Depending on the context, either a low or a high value of $t_{\mathrm{worst}}$ might be preferable. These characteristics show that our algorithm provides us with a very good estimation of the feasibility region from relatively few data points in comparison with grid-based or LHS approaches.}\label{tab:chem-results}
\begin{tabular}{|c||c|c|c|c|c|c|c|} \hline
\rule{0pt}{2.5ex} Setup & $R$ $\uparrow$ & $\sigma$ $\uparrow$ & $r_{\mathrm{fp}}$ $\downarrow$ & $r_{\mathrm{fn}}$ $\downarrow$ & $\alpha$ $\uparrow$ & $t_{\mathrm{best}}$ $\uparrow$ & $t_{\mathrm{worst}}$ \\ \hline \hline
(1) & \num{0.913 +- 0.006} & \num{1.000} & \num{0.069 +- 0.005} & \num{0.018 +- 0.003} & \num{0.739} & \num{0.841} & \num{0.588} \\ \hline
(2) & \num{0.863 +- 0.007} & \num{1.000} & \num{0.134 +- 0.006} & \num{0.003 +- 0.001} & \num{1.105} & \num{0.894} & \num{0.584} \\ \hline
(3) & \num{0.921 +- 0.005} & \num{1.000} & \num{0.047 +- 0.004} & \num{0.033 +- 0.003} & \num{0.841} & \num{0.833} & \num{0.573} \\ \hline
(4) & \num{0.724 +- 0.009} & \num{1.000} & \num{0.276 +- 0.007} & \num{0.000 +- 0.001} & \num{1.314} & \num{0.930} & \num{0.582} \\ \hline
(5) & \num{0.479 +- 0.010} & \num{1.000} & \num{0.000 +- 0.001} & \num{0.521 +- 0.007} & \num{0.066} & \num{0.725} & \num{0.725} \\ \hline
(6) & \num{0.485 +- 0.010} & \num{1.000} & \num{0.000 +- 0.001} & \num{0.515 +- 0.007} & \num{0.138} & \num{0.817} & \num{0.633} \\ \hline
$\langle$7$\rangle$ & \num{0.480 +- 0.010} & \num{0.875} & \num{0.000 +- 0.001} & \num{0.521 +- 0.007} & \num{0.481} & \num{0.847} & \num{0.567} \\ \hline
$\langle$8$\rangle$ & \num{0.521 +- 0.010} & \num{0.988} & \num{0.479 +- 0.007} & \num{0.000 +- 0.001} & \num{0.350} & \num{0.856} & \num{0.585} \\ \hline
$\langle$9$\rangle$ & \num{0.667 +- 0.009} & \num{0.984} & \num{0.333 +- 0.008} & \num{0.000 +- 0.001} & \num{0.326} & \num{0.902} & \num{0.556} \\ \hline
\end{tabular}
\end{table}

\section{Conclusions and Outlook}
From our benchmark, we have found that our method yields better exploration characteristics than a previously suggested Kriging-based approach~\cite{Boukouvala2012,Wang2017} for a binary feasibility classification scenario as soon as a critical number of sampling points has been exceeded. We have also seen that the ratio of valid to invalid data points in the sampling is much higher with our strategy, which can be important if the data set should be further used, e.\,g., to train a shortcut model or for optimization purposes. The performance of the Kriging-based approach might have been worsened by the lack of knowledge about a continuous feasibility constraint violation in our example. Therefore, we have suggested an improvement of the original approach for such cases.\par
The chemical process simulation has shown that our data exploration method provides us with an excellent approximation of the data space topology from a relatively small number of data points in comparison with grid-based or LHS approaches. By tuning the explorative hyperparameters we can intuitively control the exploration behavior. In this way we can concentrate the exploration on parameter regions with a relatively good optimization target. Hereby, we have discovered an almost perfect optimization target which could be used as a very suitable starting point for rigorous optimization strategies to speed up the optimization process. Since the simulated chemical process is fully realistic and industrially relevant we have demonstrated that our method is applicable to a real-world problem.\par
It is important to emphasize that some industrially relevant applications certainly require more than four design parameters. In such cases both our data exploration method and the previously suggested Kriging-based approach might be challenged by the exploration of a high dimensional parameter space~\cite{Wang2017}. To circumvent this curse of dimensionality, we propose to couple the data exploration strategy with a dimensionality reduction method~\cite{Maaten2008} in order to restrict the search to a lower dimensional manifold. However, an extended discussion of such an approach is beyond the scope of this manuscript.\par
Naturally, our method introduces an additional computational overhead in comparison with conventional data exploration strategies like a regular grid or a randomized sampling. Therefore, it is best suited for simulation environments where suggesting the next data point to evaluate takes significantly less time than the actual evaluation. Simulations of chemical processes constitute perfect candidates for this requirement due to their computational complexity and the difficulty of predicting their operation window.\par
As a stopping criterion for exploration we have used a fixed number of evaluations. This has allowed us to compare our method with grid-based and random sampling approaches. In practice one might instead want to make use of a suitable precision measure for the estimator to stop the evaluation at a sufficient accuracy. In the scope of our benchmark we have suggested to use the fraction of the relative success rate to the score as a possible stopping criterion comparable to a saturated exploration.\par
We have shown that the utility function strongly influences the outcome of our algorithm. It seems natural to ask in how far this function could be modified or generalized. In the following we will therefore briefly discuss open questions for future research related to this topic.\par
First of all, from a more general point of view, maximizing the utility function can also be regarded as a MCO problem, where each element of the utility vector corresponds to an objective function. In this sense the complete parameter space represents the feasible set of decision vectors. We use the weight vector to reduce this problem to a single objective problem. However, it could be insightful to treat the data exploration multicriterially. In a similar way it would also be possible to take a non-scalarized MCO target into account (e.\,g., in our application from~\cref{sec:Application} both mass fractions could act as optimization targets). As indicated in the introduction, such a scenario is common in the context of chemical process engineering and therefore of great practical interest.\par
Furthermore, we always assume a constant weight vector for the entire duration of the exploration. One possible modification of our algorithm would be a dynamic approach in which the coefficients of the weight vector change in each step of the iteration. For example, an exploration which starts with a strong weight of the outcome prediction uncertainty and ends with a strong weight of the optimization target prediction would first focus on the classification boundary and later on the region of interest. Such a dynamic approach could also be combined with a MCO strategy.\par
In this manuscript we have focused on a binary classification approach to separate valid from invalid simulation outcomes. However, for certain applications it could be beneficial to distinguish different causes of invalidity. One could for example use a ternary classification approach to separate valid, physically unfeasible and numerically divergent outcomes. Depending on the simulation, even more different invalidity classes might be of interest. Our exploration method could be straightforwardly generalized to incorporate such an extension. SVMs could still be used to calculate prediction probabilities for such a multi-class classification problem~\cite{Wu2004}. Since the task of classification itself is still an active area of research with connections to many other scientific fields (such as quantum mechanics~\cite{Schuld2015} with promising results~\cite{Bauckhage2018,Kerenidis2018}), it can be expected that the quality and performance of classification methods is subject to future improvements.\par
In real-world applications the outcomes of a simulation may crucially depend on a large number of different simulation parameters (e.\,g., algorithmic parameters or initialization values) and their mutual interactions. This means that divergent outcomes could turn into feasible outcomes for different simulation parameters and vice versa. However, in a complex simulation environment the reasons for a numerical divergence can become practically untraceable. To take this behavior into account, it would be possible to make use of a stochastic perspective, where numerically divergent outcomes are only considered invalid with a certain probability. This invalidity probability would consequently reflect the ignorance about the influence of the simulation parameters on the outcomes.\par
Prediction probabilities for the (multi-class) classification of outcomes already provide a statistical framework that can be exploited in this context. It would for example allow to identify numerically divergent data points of high uncertainty, i.\,e., data points that result from a divergent simulation run but have a comparably low probability of belonging to the class of divergent outcomes. Such data points could then be re-evaluated (and possibly re-labeled if a different outcome occurs) using differently tuned simulation parameters. If the simulation environments allows it, a higher computational effort (e.\,g., by increasing the numerical precision or the iteration steps of the underlying equation solver) could be used for each revision in the hope that convergence can eventually be achieved. The allocation of suitable sample weights~\cite{Osuna1997} would be an intrinsic way of SVMs to assign a higher certainty to re-calculated data points with the same divergent outcome.\par
Although the proposed statistical perspective fits rather naturally into our new method of optimized data exploration, we only consider it a conceptional idea that goes beyond the scope of this manuscript. Furthermore, the benefit of re-evaluating divergent data points harshly depends on the specific simulation environment and requires full control over the simulation parameters.\par
Summarized, we consider the straightforward variability of our algorithm through a modification of the utility function as a conceptional strength, which allows us to study different approaches in a single framework. Therefore, our method can also be seen as a very versatile starting point for further research in the fields of data exploration, feasibility classification and optimization. Moreover, its demonstrated practicability gives way for different kind of applications in the realm of chemical process engineering and beyond. 

\section{Acknowledgements}
This work was developed in the Fraunhofer Cluster of Excellence ``Cognitive Internet Technologies''. The authors would like to thank Christian Bauckhage, J{\"u}rgen Franke and Marius Kloft for their helpful and constructive comments. Our numerical examples were realized with the help of \emph{scikit-learn}~\cite{scikitlearn}.

\bookmarksetupnext{level=part}
\begin{appendices}

\section{Kernel methods}\label{app:kernel methods}
As explained in~\cref{sec:Framework:prediction} we use a kernel SVM and a kernel RR to perform predictions,~\cref{eqn:Emap}. Both estimators rely on the kernel method~\cite{Hastie2013}. Specifically, we assume that there exists a mapping
\begin{align} \label{eqn:phi}
\phi : \boldsymbol{\chi} \rightarrow \mathcal{F} 
\end{align}
from the parameter space $\boldsymbol{\chi}$ to a feature space $\mathcal{F}$ in such a way that the inner product
\begin{align}
\langle \phi(\mathbf{x}), \phi(\mathbf{x}^{\prime}) \rangle \equiv k(\mathbf{x},\mathbf{x}^{\prime})
\end{align}
represents a Gaussian kernel
\begin{align} \label{eqn:k}
k(\mathbf{x},\mathbf{x}^{\prime}) \equiv \exp ( - \gamma || \mathbf{x} - \mathbf{x}^{\prime} ||_2^2 )
\end{align}
for all parameters $\mathbf{x}$ and $\mathbf{x}^{\prime}$ in $\boldsymbol{\chi}$, where $\gamma$ is a hyperparameter of the feature space metric and $|| \cdot ||_2$ stands for the 2-norm distance.\par
It is important to emphasize that the feature space for the SVM is not necessarily the same as for the RR. Therefore, we assume in fact two feature space mappings, namely $\phi_{\mathrm{C}}$ for the SVM, which maps to the classification feature space $\mathcal{F}_{\mathrm{C}}$, and $\phi_{\mathrm{R}}$ for the RR, which maps to the regression feature space $\mathcal{F}_{\mathrm{R}}$. Both mappings are defined in analogy to~\cref{eqn:phi} and are associated with the Gaussian kernels $k_{\mathrm{C}}(\mathbf{x},\mathbf{x}^{\prime})$ and $k_{\mathrm{R}}(\mathbf{x},\mathbf{x}^{\prime})$, respectively, of the form given by~\cref{eqn:k} with hyperparameters $\gamma_{\mathrm{C}}$ and $\gamma_{\mathrm{R}}$, respectively.

\section{Classification feature space distance}\label{app:classification feature space distance}
The feature space distance between two parameters $\mathbf{x_1}$ and $\mathbf{x_2}$ with the common feature mapping $\phi$,~\cref{eqn:phi}, can be written as~\cite{Scholkopf2000}
\begin{align} \label{eqn:deltaphi}
\delta\phi(\mathbf{x_1},\mathbf{x_2}) & \equiv || \phi(\mathbf{x_1}) - \phi(\mathbf{x_2}) ||_2^2 \\ \nonumber
& = \frac{1}{2} \left[ k(\mathbf{x_1}, \mathbf{x_1}) + k(\mathbf{x_2}, \mathbf{x_2}) \right] - k(\mathbf{x_1}, \mathbf{x_2})
\end{align}
based on an arbitrary kernel function $k$. We make use of this expression in~\cref{sec:Framework:utility function} to define the third component of the utility vector,~\cref{eqn:u}, as the classification feature space distance
\begin{align} \label{eqn:Ur:general}
U_r(D,\mathbf{x}) & \equiv \delta\phi_{\mathrm{C}}(\mathbf{x},\mathbf{x_{\mathrm{NN}}})
\end{align}
between the parameter $\mathbf{x}$ and its nearest neighbor $\mathbf{x_{\mathrm{NN}}}$ from the data set $D_x$,~\cref{eqn:Dx}. Here, $\phi_{\mathrm{C}}$ represents the classification feature space mapping of the kernel SVM estimator described in~\cref{app:kernel methods}. The nearest neighbor is obtained from solving
\begin{align} \label{eqn:xNN:general}
\mathbf{x_{\mathrm{NN}}} & \equiv \argmin_{\mathbf{x}^{\prime} \in D_x} \left[ \delta\phi_{\mathrm{C}}(\mathbf{x},\mathbf{x}^{\prime}) \right] = \argmin_{\mathbf{x}^{\prime} \in D_x} \left[ \frac{1}{2} k_{\mathrm{C}}(\mathbf{x}^{\prime}, \mathbf{x}^{\prime}) - k_{\mathrm{C}}(\mathbf{x}, \mathbf{x}^{\prime}) \right].
\end{align}
The choice of a Gaussian kernel $k_{\mathrm{C}}(\mathbf{x}, \mathbf{x}^{\prime})$,~\cref{eqn:k}, allows us to rewrite \cref{eqn:Ur:general,eqn:xNN:general} as~\cref{eqn:Ur}.\par
By definition, $U_r(D,\mathbf{x})$ reduces the utility the closer the parameter $\mathbf{x}$ is located to its nearest neighbor $\mathbf{x_{\mathrm{NN}}}$ in the classification feature space $\mathcal{F}_{\mathrm{C}}$, i.\,e., the more similar the parameter is to its nearest neighbor in $D_x$. For neighbors with identical features, $U_r(D,\mathbf{x})$ becomes $0$ and for neighbors with fundamentally different features it can asymptotically approach $1$.

\section{Relative success rate and score}\label{app:relative success rate and score}
To quantify the quality of the exploration we make use of the relative success rate
\begin{align} \label{eqn:Rgeneral}
R(D) \equiv R(\mathcal{S},\mathcal{E},D,\boldsymbol{\chi}) \equiv \frac{V_{\mathrm{true}}(\mathcal{S},\mathcal{E},D,\boldsymbol{\chi})}{V_{\mathrm{total}}(\boldsymbol{\chi})}.
\end{align}
It contains the volume of correct outcome predictions
\begin{align} \label{eqn:Vtrue}
V_{\mathrm{true}}(\mathcal{S},\mathcal{E},D,\boldsymbol{\chi}) \equiv \int_{\boldsymbol{\chi}} \! \mathbf{1}_{\mathrm{true}}(\mathcal{S},\mathcal{E},D,\mathbf{x}) \ \mathrm d \mathbf{x}
\end{align}
with the indicator function
\begin{align}
\mathbf{1}_{\mathrm{true}}(\mathcal{S},\mathcal{E},D,\mathbf{x}) \equiv \begin{cases}
1 & \text{if} \ \hat{y}(\mathcal{E},D,\mathbf{x}) = y(\mathcal{S},\mathbf{x}) \\
0 & \text{otherwise}
\end{cases}
\end{align}
and the total volume of the parameter space 
\begin{align} \label{eqn:Vtotal}
V_{\mathrm{total}}(\boldsymbol{\chi}) \equiv \int_{\boldsymbol{\chi}} \! \mathrm d \mathbf{x}.
\end{align}
We can also write
\begin{align} \label{eqn:Rsigma}
R(\mathcal{S},\mathcal{E},D,\boldsymbol{\chi}) = \sigma(D_{\chi}(\mathcal{S}),\mathcal{E},D),
\end{align}
where the score
\begin{align} \label{eqn:sigma:general}
\sigma(D_{\mathrm{t}}) & \equiv \sigma(D_{\mathrm{t}},\mathcal{E},D) \equiv \frac{| \{ (\mathbf{x},y,t) \in D_{\mathrm{t}} \, | \, \hat{y}(\mathcal{E},D,\mathbf{x}) = y \} |}{| \{ d(\mathbf{x}) \in D_{\mathrm{t}} \} |}
\end{align}
of a test data set $D_{\mathrm{t}}$ represents the fraction of correct outcome predictions performed for all parameters $\mathbf{x} \in D_{\mathrm{t}}$. The absolute values in~\cref{eqn:sigma:general} denote a set cardinality, a notation which we will also use in the following sections. The specific test data set
\begin{align} \label{eqb:Dchi}
D_{\chi}(\mathcal{S}) \equiv  \{ (\mathbf{x},y(\mathcal{S},\mathbf{x}),t(\mathcal{S},\mathbf{x})) \ \text{for almost all} \ \mathbf{x} \in \boldsymbol{\chi} \}
\end{align}
from~\cref{eqn:Rsigma} is required to contain data points $d(\mathbf{x})$,~\cref{eqn:d}, for all or almost all parameters $\mathbf{x}$ from $\boldsymbol{\chi}$.\par
Summarized, the relative success rate $R(D) \in [0,1]$ is a quality measure for an outcome estimator taken across the whole parameter space, whereas the score $\sigma(D_{\mathrm{t}}) \in [0,1]$ may also be used for a local quality measure depending on the chosen test data set $D_{\mathrm{t}}$. A larger value of $R(D)$ and $\sigma(D_{\mathrm{t}})$ indicates an estimator of higher quality.\par
In practice, the relative success rate can be determined by evaluting the integral in~\cref{eqn:Vtrue} with the help of a standard Monte Carlo approach~\cite{Press2007}, which also provides us with an estimated error. Given the collection of evaluated parameters
\begin{align}\label{eqn:DMCx}
D^{\mathrm{MC}}_x \equiv \{\mathbf{x}_1,\dots,\mathbf{x}_{\mathcal{N}}\}
\end{align}
in the sense of~\cref{eqn:Dx}, which have been chosen randomly from $\boldsymbol{\chi}$, we can make use of~\cref{eqn:R:approx} with the approximation
\begin{align}
R^{\mathrm{MC}}(D) & \equiv R^{\mathrm{MC}}(\mathcal{S},\mathcal{E},D,\mathcal{N}) \nonumber \\
& \equiv E^{\mathrm{MC}}[\mathbf{1}_{\mathrm{true}}(\mathcal{S},\mathcal{E},D,\mathbf{x}); D^{\mathrm{MC}}_x)
\end{align}
and its estimated error
\begin{align}
\delta R^{\mathrm{MC}}(D) \equiv \delta R^{\mathrm{MC}}(\mathcal{S},\mathcal{E},D,\mathcal{N}) \equiv & S^{\mathrm{MC}}[\mathbf{1}_{\mathrm{true}}(\mathcal{S},\mathcal{E},D,\mathbf{x}); D^{\mathrm{MC}}_x),
\end{align}
respectively. Here we have introduced the Monte Carlo expectation value
\begin{align} \label{eqn:EMC}
E^{\mathrm{MC}}[f; D_x) \equiv \frac{1}{|D_x|} \sum_{\mathbf{x}_i \in D_x} \! f(\mathbf{x}_i)
\end{align}
and the Monte Carlo error estimate
\begin{align} \label{eqn:SMC:general}
S^{\mathrm{MC}}[f; D_x, \alpha_{\mathcal{P}}) \equiv \frac{|z_{\alpha_{\mathcal{P}}/2}|}{\sqrt{|D_x| (|D_x|-1)}} \sqrt{\sum_{\mathbf{x}_i \in D_x} \! \left[ f(\mathbf{x}_i) - E^{\mathrm{MC}}[f; D_x) \right]^2},
\end{align}
which can both be considered a functional with respect to a function $f: \boldsymbol{\chi} \rightarrow \mathbb{R}$. The two expressions additionally depend on a collection of evaluated parameters $D_x$,~\cref{eqn:Dx}. The symbol $z_{\alpha_{\mathcal{P}}/2}$ refers to the standard $\alpha_{\mathcal{P}}/2$ normal quantile associated with the probability $\mathcal{P} = 1-\alpha_{\mathcal{P}}$ of the precise integration result with the approximation $E^{\mathrm{MC}}[f; D_x)$ being within the confidence interval $[E^{\mathrm{MC}}[f; D_x) - S^{\mathrm{MC}}[f; D_x, \alpha_{\mathcal{P}}), E^{\mathrm{MC}}[f; D_x) + S^{\mathrm{MC}}[f; D_x, \alpha_{\mathcal{P}})]$. For all of our numerical examples we set $\mathcal{P} = 95\%$ so that $\alpha_{\mathcal{P}} = 5\%$ and $|z_{\num{.025}}| \approx \num{1.960}$. We introduce the appropriate abbreviation
\begin{align} \label{eqn:SMC}
S^{\mathrm{MC}}[f; D_x) \equiv S^{\mathrm{MC}}[f; D_x, 5\%)
\end{align}
to simplify our notation.

\section{Ratios of false positives and false negatives}\label{app:ratios of false positives and false negatives}
A common quality measure for a binary classification method is the number of false positives (for which the classification method improperly predicts a positive result) and the number of false negatives (for which the classification method improperly predicts a negative result), respectively. Following this concept, we can define the ratio of false positives
\begin{subequations}
\begin{align}
r_{\mathrm{fp}}(D_{\mathrm{t}},\mathcal{E},D) & \equiv \frac{| \{ (\mathbf{x},y,t) \in D_{\mathrm{t}} \, | \, y = \mathrm{invalid} \land \hat{y}(\mathcal{E},D,\mathbf{x}) = \mathrm{valid} \} |}{| D_{\mathrm{t}} |}
\end{align}
and the ratio of false negatives
\begin{align}
r_{\mathrm{fn}}(D_{\mathrm{t}},\mathcal{E},D) & \equiv \frac{| \{ (\mathbf{x},y,t) \in D_{\mathrm{t}} \, | \, y = \mathrm{valid} \land \hat{y}(\mathcal{E},D,\mathbf{x}) = \mathrm{invalid} \} |}{| D_{\mathrm{t}} |},
\end{align}
\end{subequations}
for the estimator $\mathcal{E}$ trained with respect to a test data set $D_{\mathrm{t}}$. By definition, a positive result corresponds to a valid outcome, whereas a negative result corresponds to an invalid outcome.\par
If we choose the specific test set $D_{\chi}(\mathcal{S})$,~\cref{eqb:Dchi}, we find
\begin{subequations} \label{eqn:r:general}
\begin{align}
r_{\mathrm{fp}}(D) & \equiv r_{\mathrm{fp}}(D_{\chi}(\mathcal{S}),\mathcal{E},D) = \frac{V_{\mathrm{fp}}(\mathcal{S},\mathcal{E},D,\boldsymbol{\chi})}{V_{\mathrm{total}}(\boldsymbol{\chi})}
\end{align}
and
\begin{align}
r_{\mathrm{fn}}(D) & \equiv r_{\mathrm{fn}}(D_{\chi}(\mathcal{S}),\mathcal{E},D) = \frac{V_{\mathrm{fn}}(\mathcal{S},\mathcal{E},D,\boldsymbol{\chi})}{V_{\mathrm{total}}(\boldsymbol{\chi})},
\end{align}
\end{subequations}
respectively, in analogy to~\cref{eqn:Rsigma}. Here we have recalled total parameter space volume $V_{\mathrm{total}}(\boldsymbol{\chi})$,~\cref{eqn:Vtotal}. Moreover, we have introduced the volume of false positives
\begin{subequations}
\begin{align}
V_{\mathrm{fp}}(\mathcal{S},\mathcal{E},D,\boldsymbol{\chi}) \equiv \int_{\boldsymbol{\chi}} \! [ 1 - \mathbf{1}_{\mathrm{true}}(\mathcal{S},\mathcal{E},D,\mathbf{x}) ] \mathbf{1}_{\mathrm{p}}(\mathcal{E},D,\mathbf{x})(\mathcal{E},D,\boldsymbol{\chi}) \ \mathrm d \mathbf{x}
\end{align}
and the volume of false negatives
\begin{align}
V_{\mathrm{fn}}(\mathcal{S},\mathcal{E},D,\boldsymbol{\chi}) \equiv \int_{\boldsymbol{\chi}} \! [ 1 - \mathbf{1}_{\mathrm{true}}(\mathcal{S},\mathcal{E},D,\mathbf{x}) ] \mathbf{1}_{\mathrm{n}}(\mathcal{E},D,\mathbf{x})(\mathcal{E},D,\boldsymbol{\chi}) \ \mathrm d \mathbf{x},
\end{align}
\end{subequations}
which are based on the indicator functions
\begin{subequations}
\begin{align}
\mathbf{1}_{\mathrm{p}}(\mathcal{E},D,\mathbf{x}) \equiv \begin{cases}
1 & \text{if} \ \hat{y}(\mathcal{E},D,\mathbf{x}) = \mathrm{valid} \\
0 & \text{otherwise}
\end{cases}
\end{align}
and
\begin{align}
\mathbf{1}_{\mathrm{n}}(\mathcal{E},D,\mathbf{x}) \equiv 1 - \mathbf{1}_{\mathrm{p}}(\mathcal{E},D,\mathbf{x}),
\end{align}
\end{subequations}
respectively.\par
Summarized, the ratios $r_{\mathrm{fp}}(D) \in [0,1]$ and $r_{\mathrm{fn}}(D) \in [0,1]$ can be considered as a quality measure for an outcome estimator taken across the whole parameter space. A worst-case reference value for the ratios of false positives and false negatives will be introduced in~\cref{app:validity ratio} based on evaluating randomized predictions for almost all parameters in $\boldsymbol{\chi}$. Note that an estimator of high quality is indicated by a large relative success rate $R(D)$,~\cref{eqn:Rgeneral}, but small ratios of false positives and false negatives; also see~\cref{eqn:rR}.\par
In complete analogy to~\cref{app:relative success rate and score}, we can use a Monte Carlo approach to numerically calculate $V_{\mathrm{fp}}(\mathcal{S},\mathcal{E},D,\boldsymbol{\chi})$ and $V_{\mathrm{fn}}(\mathcal{S},\mathcal{E},D,\boldsymbol{\chi})$ in~\cref{eqn:r:general} using a collection $D^{\mathrm{MC}}_x$ of $\mathcal{N}$ evaluated parameters,~\cref{eqn:DMCx}, which have been chosen randomly from $\boldsymbol{\chi}$. This leads us to~\cref{eqn:r:approx} with the approximations 
\begin{subequations}
\begin{align}
r_{\mathrm{fp}}^{\mathrm{MC}}(D) \equiv r_{\mathrm{fp}}^{\mathrm{MC}}(\mathcal{S},\mathcal{E},D,\mathcal{N}) \equiv E^{\mathrm{MC}}[\mathbf{1}_{\mathrm{fp}}(\mathcal{S},\mathcal{E},D,\mathbf{x}); D^{\mathrm{MC}}_x)
\end{align}
and
\begin{align}
r_{\mathrm{fn}}^{\mathrm{MC}}(D) \equiv r_{\mathrm{np}}^{\mathrm{MC}}(\mathcal{S},\mathcal{E},D,\mathcal{N}) \equiv E^{\mathrm{MC}}[\mathbf{1}_{\mathrm{fn}}(\mathcal{S},\mathcal{E},D,\mathbf{x}); D^{\mathrm{MC}}_x),
\end{align}
\end{subequations}
as well as their estimated errors
\begin{subequations}
\begin{align}
\delta r_{\mathrm{fp}}^{\mathrm{MC}}(D) \equiv \delta r_{\mathrm{fp}}^{\mathrm{MC}}(\mathcal{S},\mathcal{E},D,\mathcal{N}) \equiv & S^{\mathrm{MC}}[\mathbf{1}_{\mathrm{fp}}(\mathcal{S},\mathcal{E},D,\mathbf{x}); D^{\mathrm{MC}}_x)
\end{align}
and
\begin{align}
\delta r_{\mathrm{fn}}^{\mathrm{MC}}(D) \equiv \delta r_{\mathrm{fn}}^{\mathrm{MC}}(\mathcal{S},\mathcal{E},D,\mathcal{N}) \equiv & S^{\mathrm{MC}}[\mathbf{1}_{\mathrm{fp}}(\mathcal{S},\mathcal{E},D,\mathbf{x}); D^{\mathrm{MC}}_x),
\end{align}
\end{subequations}
respectively. Here we have recalled the Monte Carlo expectation value,~\cref{eqn:EMC}, and the Monte Carlo error estimate,~\cref{eqn:SMC}, to simplify our notation.

\section{Validity ratio}\label{app:validity ratio}
In general, valid data points are more insightful than invalid data points because they contain information about the optimization target. To quantify the usefulness of a sampling, we therefore introduce the validity ratio
\begin{align} \label{eqn:alphageneral}
\alpha(D) \equiv \frac{| \{ (\mathbf{x},y,t) \in D \, | \, y = \mathrm{valid} \} |}{| \{ (\mathbf{x},y,t) \in D \, | \, y = \mathrm{invalid} \} |}.
\end{align}
The higher the fraction $\alpha(D)$ of valid to invalid data points in the data set $D$, the more useful the corresponding sampling.\par
However, to provide a clear scale for $\alpha$, some kind of reference value is necessary. Such a reference value can be obtained from the idea of an infinite sample: For a given simulation $\mathcal{S}$, a fully dense sampling in the parameter space $\boldsymbol{\chi}$ with infinitely many data points results in the infinitely large data set $D_{\infty}(\boldsymbol{\chi})$. For this reference data set the validity rate converges to the reference limit
\begin{align}
\lim_{D \rightarrow D_{\infty}(\mathcal{S}, \boldsymbol{\chi})}{\alpha(D)} = \alpha_{\infty}(\mathcal{S}, \boldsymbol{\chi})
\end{align}
given by the fractions
\begin{align} \label{eqn:alphainfgeneral}
\alpha_{\infty} \equiv \alpha_{\infty}(\mathcal{S}, \boldsymbol{\chi}) \equiv \frac{V_{\mathrm{valid}}(\mathcal{S},\boldsymbol{\chi})}{V_{\mathrm{invalid}}(\mathcal{S},\boldsymbol{\chi})}
\end{align}
of the total volumes for valid and invalid outcomes in $\boldsymbol{\chi}$, respectively. Here we have introduced the valid volume
\begin{align} \label{eqn:Vvalid}
V_{\mathrm{valid}}(\mathcal{S},\boldsymbol{\chi}) \equiv \int_{\boldsymbol{\chi}} \! \mathbf{1}_{\mathrm{valid}}(\mathcal{S},\mathbf{x}) \ \mathrm d \mathbf{x}
\end{align}
with the indicator function
\begin{align} \label{eqn:1valid}
\mathbf{1}_{\mathrm{valid}}(\mathcal{S},\mathbf{x}) \equiv \begin{cases}
1 & \text{if} \ y(\mathcal{S},\mathbf{x}) = \mathrm{valid} \\
0 & \text{otherwise}
\end{cases}
\end{align}
and the invalid volume
\begin{align} \label{eqn:Vinvalid}
V_{\mathrm{invalid}}(\mathcal{S},\boldsymbol{\chi}) \equiv V_{\mathrm{total}}(\boldsymbol{\chi}) - V_{\mathrm{valid}}(\mathcal{S},\boldsymbol{\chi})
\end{align}
based on the total parameter space volume $V_{\mathrm{total}}(\boldsymbol{\chi})$,~\cref{eqn:Vtotal}.\par
For the two-dimensional toy example from~\cref{sec:Demonstration} the reference limit $\alpha_{\infty}(\mathcal{S}_{\mathrm{toy}}, \boldsymbol{\chi}_{\mathrm{toy}})$,~\cref{eqn:alphainfgeneral}, can be calculated explicitly using the definitions of the parameter space $\boldsymbol{\chi}_{\mathrm{toy}}$ and the toy simulation $\mathcal{S}_{\mathrm{toy}}$,~\cref{eqn:toy:chi,eqn:toy}, respectively. It is straightforward to see from~\cref{fig:toy-pspace} that the valid volume 
\begin{align}
V_{\mathrm{valid}}(\mathcal{S}_{\mathrm{toy}},\boldsymbol{\chi}_{\mathrm{toy}}) = 2 \frac{(\sqrt{3})^2 \pi}{4} = \frac{3 \pi}{2}
\end{align}
corresponds to the area of two quarters of a circle of radius $\sqrt{3}$. On the other hand, the total parameter space volume 
\begin{align}
V_{\mathrm{total}}(\boldsymbol{\chi}_{\mathrm{toy}}) = 4^2 = 16
\end{align}
is equivalent to the area of a square of side length $4$. According to~\cref{eqn:Vinvalid} the invalid volume 
\begin{align}
V_{\mathrm{invalid}}(\mathcal{S}_{\mathrm{toy}},\boldsymbol{\chi}_{\mathrm{toy}}) = 16 - \frac{3 \pi}{2}
\end{align}
is the difference of the two aforementioned volumes. Thus, the reference limit
\begin{align} \label{eqn:alphainf:toy:derivation}
\alpha_{\infty}(\mathcal{S}_{\mathrm{toy}}, \boldsymbol{\chi}_{\mathrm{toy}}) = \frac{ \frac{3 \pi}{2}}{16 - \frac{3 \pi}{2}} = \frac{3 \pi}{32 - 3 \pi}
\end{align}
is given by~\cref{eqn:alphainf:toy}.\par
For the chemical process simulation from~\cref{sec:Application}, an explicit expression of the reference limit of the validity rate can not be obtained since we use a flowsheet simulator to perform calculations. Therefore, we have to fall back to a numerical approximation using a Monte Carlo approach,~\cref{eqn:alphainf:approx}. We will discuss this method further below in more detail.\par
Our previous considerations also allow us to define a reference value for the ratio of false positives and false negatives,~\cref{eqn:r:general}, from~\cref{app:ratios of false positives and false negatives} in a straightforward way. Suppose a completely randomized estimator, who predicts valid and invalid outcomes with the same chance. For a fully dense sampling $D_{\infty}(\boldsymbol{\chi})$, the ratios with respect to this randomized estimator converge to the worst-case reference limits
\begin{subequations}
\begin{align}
\lim_{D \rightarrow D_{\infty}(\mathcal{S}, \boldsymbol{\chi})}{r_{\mathrm{fp}}(D)} = r_{\mathrm{fp} \infty}(\mathcal{S}, \boldsymbol{\chi})
\end{align}
and
\begin{align}
\lim_{D \rightarrow D_{\infty}(\mathcal{S}, \boldsymbol{\chi})}{r_{\mathrm{fn}}(D)} = r_{\mathrm{fn} \infty}(\mathcal{S}, \boldsymbol{\chi}),
\end{align}
\end{subequations}
respectively. One has
\begin{subequations} \label{eqn:rinfgeneral}
\begin{align}
r_{\mathrm{fp} \infty} \equiv r_{\mathrm{fp} \infty}(\mathcal{S}, \boldsymbol{\chi}) \equiv \frac{ V_{\mathrm{valid}}(\mathcal{S},\boldsymbol{\chi})}{2[V_{\mathrm{valid}}(\mathcal{S},\boldsymbol{\chi}) + V_{\mathrm{invalid}}(\mathcal{S},\boldsymbol{\chi})]} = \frac{\alpha_{\infty}}{2(\alpha_{\infty} + 1)}
\end{align}
and
\begin{align}
r_{\mathrm{fn} \infty} \equiv r_{\mathrm{fn} \infty}(\mathcal{S}, \boldsymbol{\chi}) \equiv \frac{ V_{\mathrm{invalid}}(\mathcal{S},\boldsymbol{\chi})}{2[V_{\mathrm{valid}}(\mathcal{S},\boldsymbol{\chi}) + V_{\mathrm{invalid}}(\mathcal{S},\boldsymbol{\chi})]} = \frac{1}{2(\alpha_{\infty} + 1)}
\end{align}
\end{subequations}
so that $r_{\mathrm{fp} \infty} + r_{\mathrm{fn} \infty} = \frac{1}{2}$ in general and $r_{\mathrm{fp} \infty} = r_{\mathrm{fn} \infty} = \frac{1}{4}$ for $\alpha_{\infty} = 1$, as expected for random outcome predictions.\par
For the toy example from~\cref{sec:Demonstration}, these terms can be evaluated using the explicit expression of $\alpha_{\infty}$,~\cref{eqn:alphainf:toy:derivation}, which results in~\cref{eqn:rinf:toy}. In contrast, we only have an approximation of $\alpha_{\infty}$,~\cref{eqn:alphainf:approx}, for the chemical process simulation from~\cref{sec:Application} and can therefore only use the approximations given by~\cref{eqn:rinf:approx}. This approach is explained in the following.\par
Analogously to the considerations from~\cref{app:relative success rate and score,app:ratios of false positives and false negatives}, a Monte Carlo approach allows a numerical evaluation of $V_{\mathrm{valid}}(\mathcal{S},\boldsymbol{\chi})$ in~\cref{eqn:alphainfgeneral} when we presume a collection $D^{\mathrm{MC}}_x$ of $\mathcal{N}$ evaluated parameters,~\cref{eqn:DMCx}, which have been chosen randomly from $\boldsymbol{\chi}$. As a result, we find~\cref{eqn:alphainf:approx} with the approximation
\begin{align} \label{eqn:alphainf:MC}
\alpha_{\infty}^{\mathrm{MC}} \equiv \alpha_{\infty}^{\mathrm{MC}}(\mathcal{S}, \boldsymbol{\chi},\mathcal{N}) \equiv \frac{\alpha_V(\mathcal{S},\mathcal{N})}{V_{\mathrm{total}}(\boldsymbol{\chi}) - \alpha_V(\mathcal{S},\mathcal{N})}
\end{align}
and its estimated error
\begin{align} \label{eqn:alphainf:dMC}
\delta \alpha_{\infty}^{\mathrm{MC}} \equiv \delta \alpha_{\infty}^{\mathrm{MC}}(\mathcal{S}, \boldsymbol{\chi},\mathcal{N}) \equiv \frac{V_{\mathrm{total}} \delta \alpha_V(\mathcal{S},\mathcal{N})}{[V_{\mathrm{total}}(\boldsymbol{\chi}) - \alpha_V(\mathcal{S},\mathcal{N})]^2},
\end{align}
respectively. Here we have made use of the abbreviations
\begin{align}
\alpha_V(\mathcal{S},\mathcal{N}) \equiv V_{\mathrm{total}} E^{\mathrm{MC}}[\mathbf{1}_{\mathrm{valid}}(\mathcal{S},\mathbf{x}); D^{\mathrm{MC}}_x)
\end{align}
and
\begin{align}
\delta \alpha_V(\mathcal{S},\mathcal{N}) \equiv & V_{\mathrm{total}} S^{\mathrm{MC}}[\mathbf{1}_{\mathrm{valid}}; D^{\mathrm{MC}}_x),
\end{align}
which are based on the Monte Carlo expectation value and the Monte Carlo error estimate,~\cref{eqn:EMC,eqn:SMC}, respectively.\par
Using these results, we can also numerically evaluate~\cref{eqn:rinfgeneral}. Specifically, one has~\cref{eqn:rinf:approx} with an approximation only depending on~\cref{eqn:alphainf:MC}. The respective estimated error is given by
\begin{align} \label{eqn:rinf:approx:dMC}
\delta r_{\infty}^{\mathrm{MC}} \equiv \frac{\delta \alpha_{\infty}^{\mathrm{MC}}}{2(\alpha_{\infty}^{\mathrm{MC}} + 1)^2}
\end{align}
and depends on~\cref{eqn:alphainf:MC,eqn:alphainf:dMC}, respectively.\par
Summarized, for the chemical process simulation from~\cref{sec:Application} the evaluation of~\cref{eqn:alphainf:approx} based on~\cref{eqn:alphainf:MC,eqn:alphainf:dMC} leads us to~\cref{eqn:alphainf:approx:evaluated}. Moreover, from the evaluation of~\cref{eqn:rinf:approx,eqn:rinf:approx:dMC} we find~\cref{eqn:rinf:approx:evaluated}.

\end{appendices}
%%%%%%%%%%%%%%%%%%%%%%%%%%%%%%%%%%%%%%%%%%%%%%%%%%%%%%%%%%%

\bookmarksetup{startatroot}
\addcontentsline{toc}{section}{\refname}
\bibliographystyle{custom}
\bibliography{literature}

\begin{thebibliography}{10}

\bibitem{Banerjee2010}
I.~Banerjee, S.~Pal, and S.~Maiti.
\newblock Computationally efficient black-box modeling for feasibility
  analysis.
\newblock {\em Computers \& Chemical Engineering} {\bf 34}, 1515--1521, 2010.

\bibitem{Bauckhage2018}
Christian Bauckhage, E.~Brito, K.~Cvejoski, C.~Ojeda, Rafet Sifa, and
  S.~Wrobel.
\newblock Ising models for binary clustering via adiabatic quantum computing.
\newblock In Marcello Pelillo and Edwin Hancock, editors, {\em Energy
  Minimization Methods in Computer Vision and Pattern Recognition}, pages
  3--17, Cham, 2018. Springer International Publishing.

\bibitem{Biegler1997}
L.~T. Biegler, I.~E. Grossmann, and A.~W. Westerberg.
\newblock {\em Systematic methods for chemical process design}.
\newblock Prentice Hall, Old Tappan, NJ (United States), Dec 1997.

\bibitem{Bortz2014}
M.~Bortz, J.~Burger, N.~Asprion, S.~Blagov, R.~B\"ottcher, U.~Nowak,
  A.~Scheithauer, R.~Welke, K.-H. K\"ufer, and H.~Hasse.
\newblock Multi-criteria optimization in chemical process design and decision
  support by navigation on {P}areto sets.
\newblock {\em Computers \& Chemical Engineering} {\bf 60}, 354--63, 2014.

\bibitem{Boukouvala2012}
F.~Boukouvala and M.~G. Ierapetritou.
\newblock Feasibility analysis of black-box processes using an adaptive
  sampling {K}riging-based method.
\newblock {\em Computers \& Chemical Engineering} {\bf 36}, 358--368, 2012.

\bibitem{Boukouvala2014}
F.~Boukouvala and M.~G. Ierapetritou.
\newblock Derivative-free optimization for expensive constrained problems using
  a novel expected improvement objective function.
\newblock {\em AIChE Journal} {\bf 60}, 2462--2474, 2014.

\bibitem{Burger2014}
J.~Burger, N.~Asprion, S.~Blagov, R.~B\"ottcher, U.~Nowak, M.~Bortz, R.~Welke,
  K.~K\"ufer, and H.~Hasse.
\newblock {M}ulti-{O}bjective {O}ptimization and {D}ecision {S}upport in
  {P}rocess {E}ngineering - {I}mplementation and {A}pplication.
\newblock {\em Chemie Ingenieur Technik} {\bf 86}, 1065--1072, 2014.

\bibitem{Byrd1995}
Richard~H Byrd, Peihuang Lu, Jorge Nocedal, and Ciyou Zhu.
\newblock A limited memory algorithm for bound constrained optimization.
\newblock {\em SIAM Journal on Scientific Computing} {\bf 16}, 1190--1208,
  1995.

\bibitem{Cortes1995}
C.~Cortes and V.~Vapnik.
\newblock Support-vector networks.
\newblock {\em Mach. Learn.} {\bf 20}, 273--297, September 1995.

\bibitem{Geoffrion1968}
A.~M. Geoffrion.
\newblock Proper efficiency and the theory of vector maximization.
\newblock {\em Journal of Mathematical Analysis and Applications} {\bf 22},
  618--630, 1968.

\bibitem{Hastie2013}
T.~Hastie, R.~Tibshirani, and J.~Friedman.
\newblock {\em The Elements of Statistical Learning: Data Mining, Inference,
  and Prediction}.
\newblock Springer Series in Statistics. Springer New York, New York, NY, USA,
  2013.

\bibitem{Jeong2012}
S.~Jeong, D.~Choi, and M.~Jeong.
\newblock {Feasibility Classification of New Design Points Using Support Vector
  Machine Trained by Reduced Dataset}.
\newblock {\em International Journal of Precision Engineering and
  Manufacturing} {\bf 13}, 739--746, May 2012.

\bibitem{Kerenidis2018}
Iordanis Kerenidis and Alessandro Luongo.
\newblock Quantum classification of the {MNIST} dataset via slow feature
  analysis.
\newblock {\em arXiv} 2018.

\bibitem{Matern2013}
B.~Matern.
\newblock {\em Spatial Variation}.
\newblock Lecture Notes in Statistics. Springer, New York, 2013.

\bibitem{McKay1979}
M.~D. McKay, R.~J. Beckman, and W.~J. Conover.
\newblock A comparison of three methods for selecting values of input variables
  in the analysis of output from a computer code.
\newblock {\em Technometrics} {\bf 21}, 239--245, 1979.

\bibitem{Miettinen1999}
K.~Miettinen.
\newblock {\em Nonlinear Multiobjective Optimization}.
\newblock International Series in Operations Research \& Management Science.
  Kluwer Academic Publishers, Dordrecht, NL, 2004.

\bibitem{Murphy2012}
K.P. Murphy and F.~Bach.
\newblock {\em Machine Learning: A Probabilistic Perspective}.
\newblock Adaptive Computation and Machi. MIT Press, 2012.

\bibitem{Osuna1997}
Edgar Osuna, Robert Freund, and Federico Girosi.
\newblock Support vector machines: Training and applications.
\newblock Technical report, Massachusetts Institute of Technology, 1997.

\bibitem{scikitlearn}
F.~Pedregosa, G.~Varoquaux, A.~Gramfort, V.~Michel, B.~Thirion, O.~Grisel,
  M.~Blondel, P.~Prettenhofer, R.~Weiss, V.~Dubourg, J.~Vanderplas, A.~Passos,
  D.~Cournapeau, M.~Brucher, M.~Perrot, and E.~Duchesnay.
\newblock Scikit-learn: {M}achine {L}earning in {P}ython.
\newblock {\em Journal of Machine Learning Research} {\bf 12}, 2825--2830,
  2011.

\bibitem{Platt1999}
J.~C. Platt.
\newblock Probabilistic outputs for support vector machines and comparisons to
  regularized likelihood methods.
\newblock In {\em Advances in large margin classifiers}, pages 61--74,
  Cambridge, MA, USA, 1999. MIT Press.

\bibitem{Press2007}
W.~H. Press.
\newblock {\em Numerical Recipes 3rd Edition: The Art of Scientific Computing}.
\newblock Cambridge University Press, Cambdrige, GB, 2007.

\bibitem{Rasmussen2006}
C.E. Rasmussen and C.K.I Williams.
\newblock {\em Gaussian {P}rocesses for {M}achine {L}earning}.
\newblock MIT Press, 2006.

\bibitem{Renon1968}
Henri Renon and J.~M. Prausnitz.
\newblock Local compositions in thermodynamic excess functions for liquid
  mixtures.
\newblock {\em AIChE Journal} {\bf 14}, 135--144, 1968.

\bibitem{Schittkowski2011}
K.~Schittkowski.
\newblock {\em MISQP: A Fortran Subroutine of a Trust Region SQP Algorithm for
  Mixed-Integer Nonlinear Programming}.
\newblock Department of Computer Science, University of Bayreuth, Bayreuth,
  GER, 2011.

\bibitem{Scholkopf2000}
Bernhard Sch\"{o}lkopf.
\newblock The kernel trick for distances.
\newblock In {\em Proceedings of the 13th International Conference on Neural
  Information Processing Systems}, NIPS'00, pages 283--289, Cambridge, MA, USA,
  2000. MIT Press.

\bibitem{Schuld2015}
Maria Schuld, Ilya Sinayskiy, and Francesco Petruccione.
\newblock An introduction to quantum machine learning.
\newblock {\em Contemporary Physics} {\bf 56}, 172--185, 2015.

\bibitem{Shannon1948}
C.~E. Shannon.
\newblock A mathematical theory of communication.
\newblock {\em The Bell System Technical Journal} {\bf 27}, 379--423, July
  1948.

\bibitem{Storn1997}
R.~Storn and K.~Price.
\newblock Differential evolution -- a simple and efficient heuristic for global
  optimization over continuous spaces.
\newblock {\em Journal of Global Optimization} {\bf 11}, 341--359, Dec 1997.

\bibitem{Maaten2008}
L.J.P. van~der Maaten, E.~O. Postma, and H.~J. van~den Herik.
\newblock Dimensionality reduction: A comparative review.
\newblock Technical report, Tilburg University, 2009.

\bibitem{Wang2017}
Z.~Wang and M.~Ierapetritou.
\newblock {A Novel Feasibility Analysis Method for Black-Box Processes Using a
  Radial Basis Function Adaptive Sampling Approach}.
\newblock {\em AIChE Journal} {\bf 63}, 532--550, 2017.

\bibitem{Wu2004}
T.~Wu, C.~Lin, and R.~C. Weng.
\newblock Probability estimates for multi-class classification by pairwise
  coupling.
\newblock {\em J. Mach. Learn. Res.} {\bf 5}, 975--1005, December 2004.

\end{thebibliography}

\end{document}